\documentclass[11pt]{article}

\usepackage{pifont}
\usepackage[utf8]{inputenc}
\usepackage{graphicx}
\usepackage{tikz}
\usepackage{pgfplots}
\usepackage{pgfopts}
\usepackage{rotating}
\usepackage{float}
\usepackage{xcolor}
\usepackage{colortbl}
\usepackage{longtable}
\usepackage{indentfirst}
\usepackage{fancyvrb}
\usepackage{mdframed}
\usepackage{blindtext}
\usepackage{amsmath}
\usepackage{tkz-tab}
\usepackage{adjustbox}
\usepackage{amssymb}
\usepackage{amsbsy}
\usepackage{booktabs}
\usepackage{caption}
\usepackage{parskip}
\usepackage{algorithm}
\usepackage{algpseudocode}
\usepackage{listings}
\usepackage{enumitem}
\usepackage{dirtytalk}
\usepackage{tcolorbox}
\usepackage{multirow}
\usepackage[hyphens]{url}
\usepackage{hyperref}
\usepackage{placeins}
\usepackage{verbatim}
\usepackage{pdfpages}
\usepackage{pifont}

\hypersetup{
    hidelinks,
    breaklinks
}

\usetikzlibrary{shapes, arrows, positioning, calc, fit, backgrounds}
\usetikzlibrary{shapes.geometric}
\usetikzlibrary{shapes.geometric, arrows, positioning}

\tikzset{
  decision/.style = {diamond, draw, fill=blue!10,
                     text badly centered, inner sep=4pt},
  action/.style   = {rectangle, draw, rounded corners, fill=gray!10,
                     text centered, inner sep=4pt},
  arrow/.style    = {->, thick}
}

\tcbuselibrary{skins}
\tcbuselibrary{breakable}
\tcbset{
  width=\textwidth,
  center,
  colback=blue!5!white
}

\newcommand{\ourapproach}{\textsc{InvestorNerd}\xspace }

\title{\ourapproach: An Investment and Financial Insights System Based on User Profiles}

\author{
John Castillo\textsuperscript{1} \\
\texttt{john.castillo\_tacuri@tufts.edu}
\and
Rishika Gautam\textsuperscript{2} \\
\texttt{rg4929@nyu.edu}
\and
Xinyu Wang\textsuperscript{2} \\
\texttt{xw2875@nyu.edu}
\and
Harsh Kashyap\textsuperscript{3} \\
\texttt{hkashyap\_be19@thapar.edu}
\and
Dennis Shasha\textsuperscript{2}\thanks{Corresponding author}\\
\texttt{shasha@cs.nyu.edu}\\
\textsuperscript{1}Department of Computer Science, Tufts University,\\
Medford, MA 02155, USA\\
\textsuperscript{2}Department of Computer Science, Courant Institute of Mathematical Sciences,\\
New York University, New York, NY 10012, USA \\
\textsuperscript{3}Thapar Institute of Engineering and Technology\\
}

\date{March 2026}

\begin{document}

\maketitle

% \abstract{\ourapproach is a large language model-based system that allows users to submit questions about diet and nutrition. It both (i) provides concise summaries of the research with citations and (ii) assesses the statistics and biases of the cited research. This paper outlines the system's workflow, details the prompts used, and describes the system implementation. The paper than presents accuracy results based on a comparison with systematic surveys and a comparison with state-of-the-art systems based on expert reviewers. Among the highest-evaluated of these systems, \ourapproach is unique in offering both a safety and a sophisticated source article analysis. % and a second comparison with those same systems regarding the tendency to hallucinate. 
% \ourapproach can be accessed at \url{https://dietnerd.org/} and we will make the code available upon publication.}

\begin{abstract}
\ourapproach is a web-based platform (\url{investornerd.org}) designed to educate and democratize financial understanding by providing accessible, AI-powered investment and personal finance insights tailored to potential user profiles. The system addresses a key challenge in helping everyday individuals, especially those without formal financial education, make sense of investment options and personal financial decisions. The platform offers three interactive tools:
\begin{enumerate}
  \item \textbf{Stock Insights} - Allows users to input any Stock, Mutual Fund, or Exchange Traded Fund (ETF) ticker to receive a summary of relevant news sentiment and quantitative metrics.
  \item \textbf{Stock Insights Questionnaire} - Builds on stock insights by letting users specify a preferred risk tolerance and sector interest. It returns categorized investment tables based on volatility and sector, sortable by dividend yield, return percentages, and others.
  \item \textbf{General Insights Questionnaire} - Provides users with personalized insights based on answers to an age-income-expenditure-savings questionnaire. Outputs include possible actions regarding savings strategies, account types (e.g., Roth IRA, UTMA), and potential loan options (e.g., FHA).
\end{enumerate}
The output is designed to be clear, unbiased, and educational, empowering novice investors with practical insights. This paper details the design, implementation, and evaluation of \ourapproach, demonstrating how generative AI and open financial data can be integrated to create scalable, insight-rich tools for financial literacy.

\noindent\textbf{Video abstract:} \url{https://www.dropbox.com/scl/fi/zc30nznq3vjw0p3az630s/2025-12-07investornerd_abstract.mov?rlkey=aop3tiwp419bdrwc2lpg6plbt&st=6oqdyetp&dl=0}
\end{abstract}

\noindent\textbf{Keywords:} financial literacy; large language models; generative AI; question-answering; personal investment; stocks; mutual funds; ETFs

\section{Introduction}

While financial education has become increasingly accessible through platforms like \href{https://www.investopedia.com/personal-finance}{Investopedia}
, \href{https://www.ramseysolutions.com/}{Ramsey
 Solutions}, and other online financial literacy websites, general financial planning or interpreting investment vehicles (e.g., stocks) remains a complex and fragmented task. Novice public market investors often find a gap between understanding basic investment principles and applying that knowledge to evaluate specific companies or market sectors. Although raw financial data, earnings reports, and news headlines are readily available, synthesizing this information into coherent, actionable insights  demands time and expertise.

\ourapproach addresses this gap by providing real-time insights on Stocks, Mutual Funds, and ETFs in an accessible format. \ourapproach accesses financial data through \href{https://ranaroussi.github.io/yfinance/index.html}{YFinance}
 and applies a large language model-based analysis to generate educational overviews regarding investment possibilities.

The platform is designed around three primary use cases:
\begin{enumerate}
    \item \textbf{Individual Stock Exploration} – Users can search for a stock to review publicly available news and high-level educational summaries that highlight both qualitative and  quantitative factors relevant to the company.
    \item \textbf{Sector-Based Discovery} – Users can explore one or more sectors (e.g. technology and finance) to view a list of widely tracked companies separated by risk class and then obtain an analysis of one or more individual stocks.  
    \item \textbf{Financial Literacy Questionnaire} – Users may complete a short questionnaire that, when processed, provides general educational material. These outputs are designed to improve financial literacy, suggest broad areas of consideration, and help users generate ideas for their own independent research.
\end{enumerate}

Unlike tools designed for high-frequency trading, portfolio optimization, or algorithmic stock selection, \ourapproach is built explicitly to   help novice to intermediate investors, students, and individuals who want to understand financial markets. The platform provides overviews of stock performances, risk profiles, and sector trends (e.g., for technology, financial services, and utilities) grounded in both financial metrics and summarized market commentary.

This paper first shows how a user navigates the (live) system, then presents the full-stack implementation of \ourapproach. That implementation includes its user interface design, backend architecture, large language model prompt engineering strategies, and the use of two external tools: (\href{https://openai.com/}{OpenAI} and \href{https://www.perplexity.ai/hub/getting-started}{Perplexity AI}). 

The contributions of this work are:
\begin{itemize}
\item 
A working system at \href{https://www.investornerd.org}{InvestorNerd} that can be accessed for free by the general public for informational and educational purposes.
\item 
A description of the algorithms, software architecture, prompts, and interface of a system designed to provide accessible  content about stock market data and general financial literacy topics.
%\item 
%Risk-tracking experiments to examine whether \ourapproach, when using historical market data (e.g., from a previous six-month or one-year period $y$), can, with high correlation, predict the risk profile for the subsequent period $y+1$.
\item 
A comparison with state-of-the-art investment insight tools with respect to functionality, content, and user appreciation.
%Exploratory feedback from finance professionals on the potential usefulness of the system as a supplemental educational tool (not as a source of investment advice).

\end{itemize}

\section{\ourapproach in Action}

\begin{figure}[!htbp]
\centering
%\centering %% If there is a figure in wide page, please release command \centering
\begin{tikzpicture}[node distance=0.7cm and 0.7cm]
% ---- Styles ----
\tikzstyle{screenshot}=[
  rectangle, draw, thick,
  inner sep=0pt,           % no extra padding around images
  minimum width=0pt,       % let the image set the size
  minimum height=0pt,
  align=center
]
\tikzstyle{annotation}=[text width=5.5cm, align=left, font=\footnotesize]
% Screenshots - First row
\node (img1) [screenshot] at (0,0) {\includegraphics[width=0.95\textwidth]{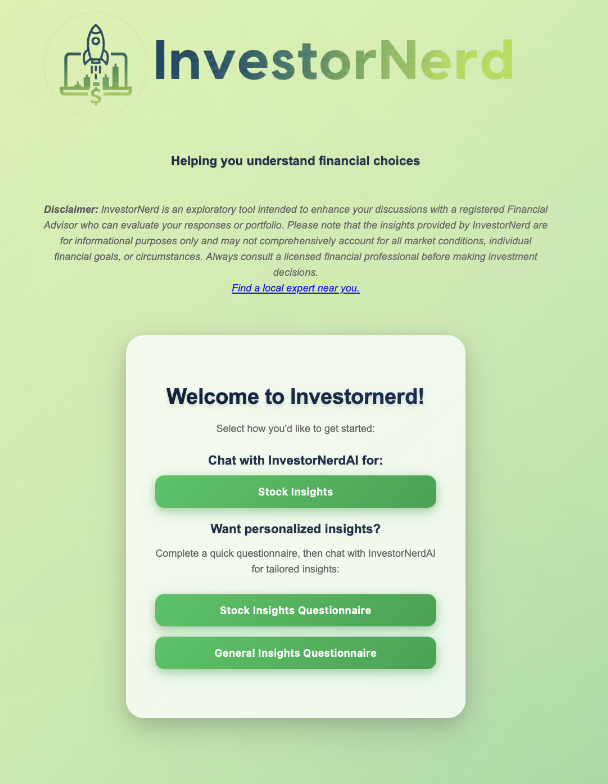}};
\end{tikzpicture}
\caption{\ourapproach landing page: the user can choose between \newline Stock Insights, Stock Insights Questionnaire, or General Insights \newline Questionnaire.}
\label{fig:landing-page-fig}
\end{figure}

\ourapproach can be accessed at \url{https://www.investornerd.org/}. 
When users open the landing page, as shown in Figure~\ref{fig:landing-page-fig},  they are prompted to choose between Stock Insights, Stock Insights Questionnaire, or General Insights Questionnaire.

\begin{figure}[!htbp]
\centering
%\centering %% If there is a figure in wide page, please release command \centering
\begin{tikzpicture}[node distance=0.7cm and 0.7cm]
% ---- Styles ----
\tikzstyle{screenshot}=[
  rectangle, draw, thick,
  inner sep=0pt,           % no extra padding around images
  minimum width=0pt,       % let the image set the size
  minimum height=0pt,
  align=center
]
\tikzstyle{annotation}=[text width=5.5cm, align=left, font=\footnotesize]
% Screenshots - First row
\node (img1) [screenshot] at (0,0) {\includegraphics[width=0.99\textwidth]{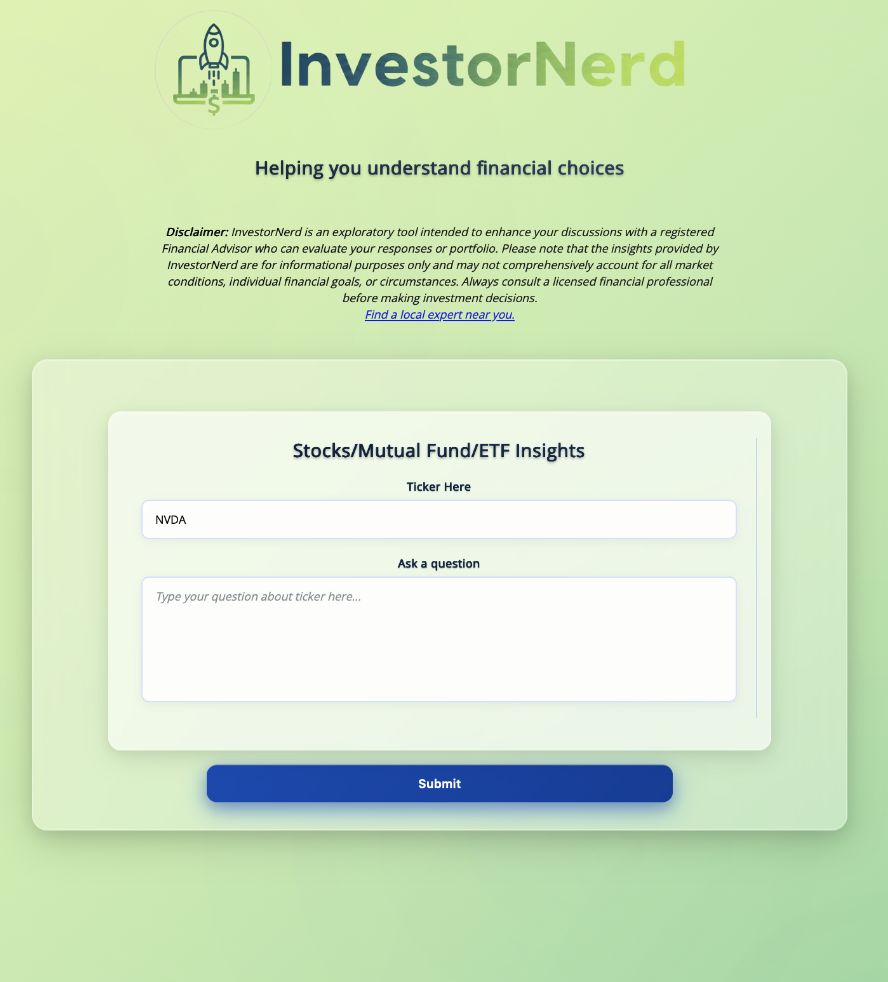}};
\end{tikzpicture}
\caption{Stock insight: the user inputs the stock ticker NVDA}
\label{fig:inputting1-fig}
\end{figure}

\begin{figure}[!htbp]
\centering
%\centering %% If there is a figure in wide page, please release command \centering
\begin{tikzpicture}[node distance=0.7cm and 0.7cm]
% ---- Styles ----
\tikzstyle{screenshot}=[
  rectangle, draw, thick,
  inner sep=0pt,           % no extra padding around images
  minimum width=0pt,       % let the image set the size
  minimum height=0pt,
  align=center
]
\tikzstyle{annotation}=[text width=5.5cm, align=left, font=\footnotesize]
% Screenshots - First row
\node (img1) [screenshot] at (0,0) {\includegraphics[width=0.95\textwidth]{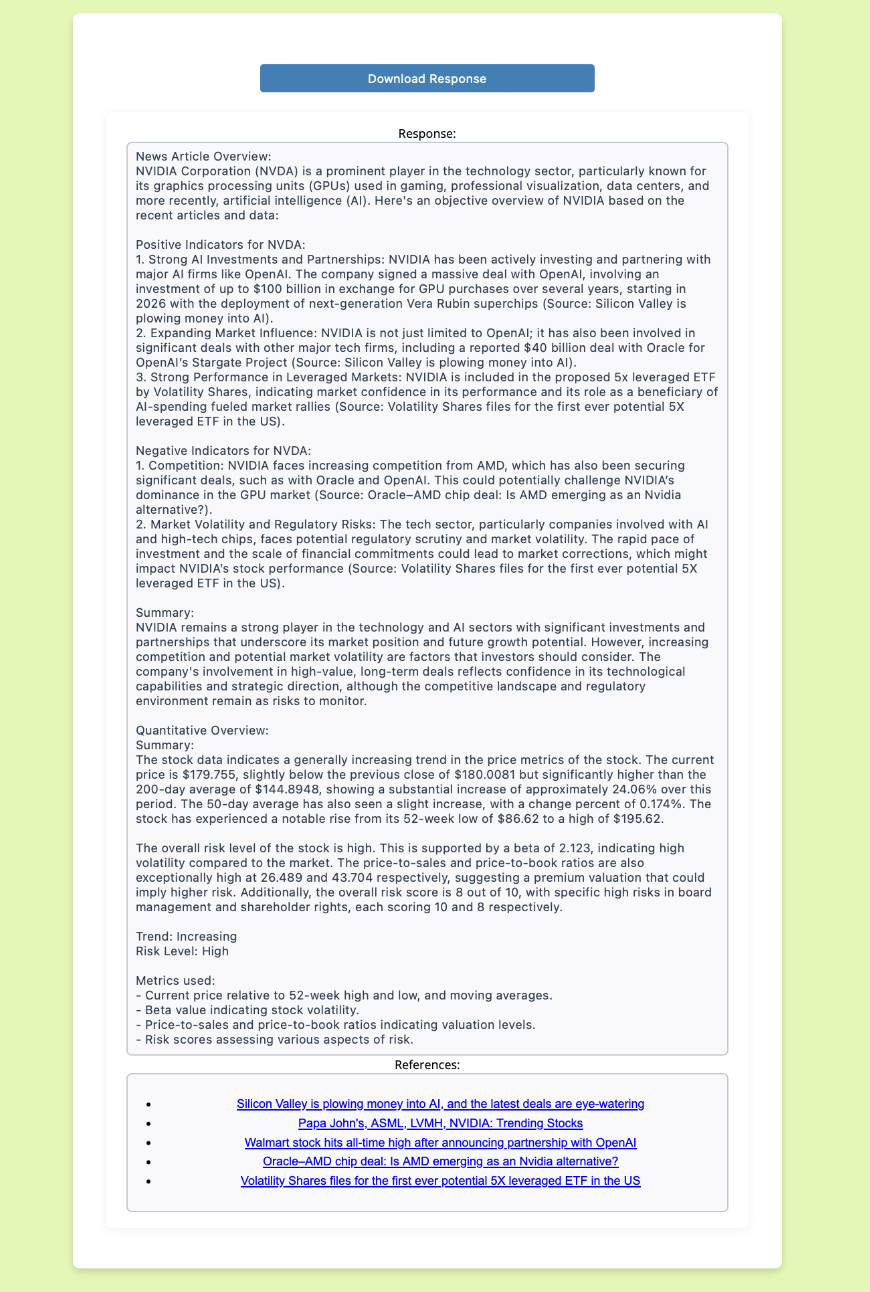}};
\end{tikzpicture}
\caption{Stock insights results: analysis generated after the user chooses Nvidia. }
\label{fig:Nvidia-answer-fig}
\end{figure}

If users have a specific stock in mind and select Stock Insight, they are prompted to enter a stock ticker, along with an optional question related to that stock. Figure~\ref{fig:inputting1-fig} illustrates the landing page with Nvidia entered. The user gets the same answer as they would have had when asking for Nvidia from the stock insights questionnaire, i.e. from Figure~\ref{fig:Nvidia-answer-fig}. % the Figure~\ref{fig:inputting2-fig} shows the processing page, and Figure~\ref{fig:inputting3-fig} presents the corresponding analysis generated by \ourapproach.

\begin{figure}[!htbp]
\centering
%\centering %% If there is a figure in wide page, please release command \centering
\begin{tikzpicture}[node distance=0.7cm and 0.7cm]
% ---- Styles ----
\tikzstyle{screenshot}=[
  rectangle, draw, thick,
  inner sep=0pt,           % no extra padding around images
  minimum width=0pt,       % let the image set the size
  minimum height=0pt,
  align=center
]
\tikzstyle{annotation}=[text width=5.5cm, align=left, font=\footnotesize]
% Screenshots - First row
\node (img1) [screenshot] at (0,0) {\includegraphics[width=0.97\textwidth]{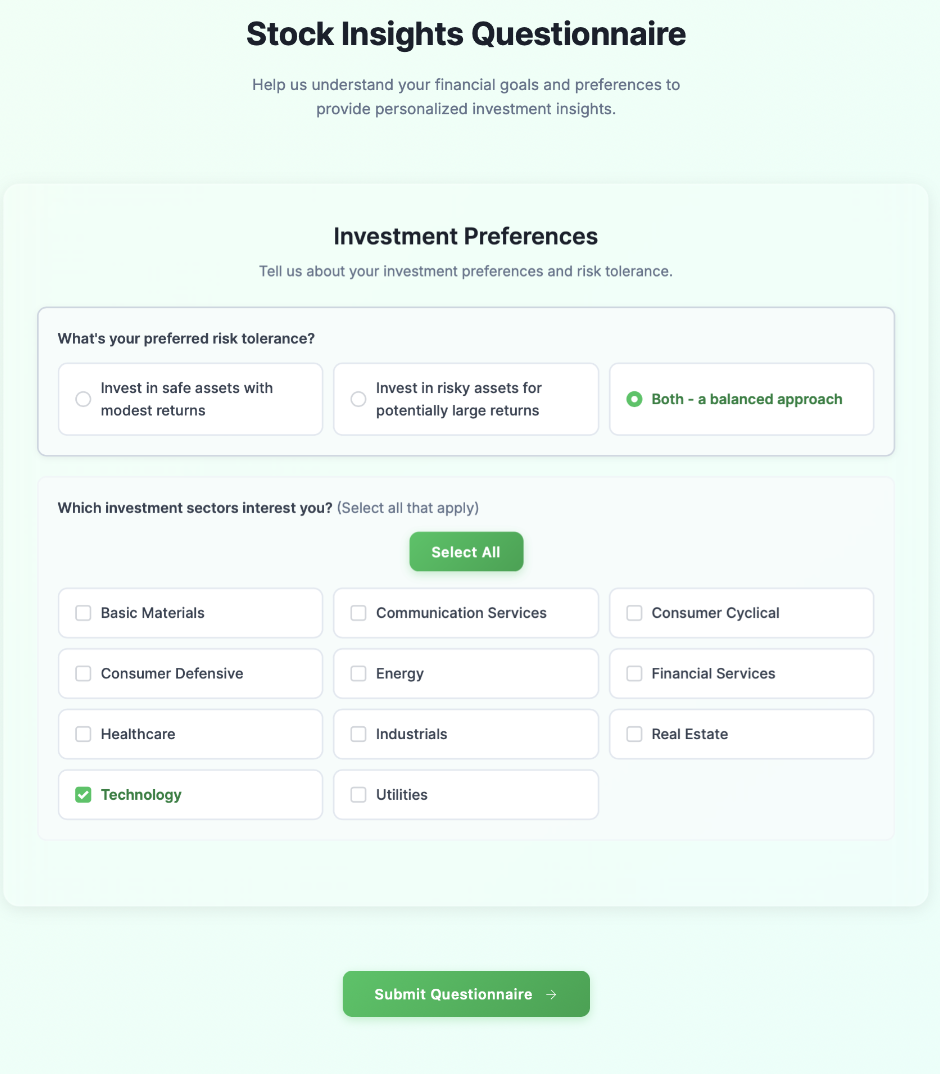}};
\end{tikzpicture}
\caption{Answered questionnaire: this is a completed questionnaire for a user  interested in the technology sector at different risk levels.}
\label{fig:stock-answers1-fig}
\end{figure}

If the user selects the Stock Insights Questionnaire, they are prompted to complete a form %(Figure~\ref{fig:stock-quests-fig}), 
where they specify their risk tolerance and sectors of interest. Figure~\ref{fig:stock-answers1-fig} shows a completed questionnaire for a user who is willing to take risks and is interested in the Technology sector. Based on these inputs, \ourapproach generates a spreadsheet (Figure~\ref{fig:stock-results1-fig} and Figure~\ref{fig:stock-results2-fig}) containing stocks that have historically met the user's risk profile and sector preference. %Figure~\ref{fig:Nvidia-answer-fig} displays the analysis provided after the user selects Nvidia, as an example. Figure~\ref{fig:stock-answers2-fig} presents a questionnaire completed by a user who is aggressive and interested in the Communication Services sector. 
Panels Figure~\ref{fig:stock-results3-fig} and Figure~\ref{fig:stock-results4-fig} show the resulting spreadsheet tailored to this profile. Figure~\ref{fig:EA-answer-fig} displays the analysis generated after selecting EA, an aggressive stock. EA was determined an aggressive stock based on how InvestorNerd calculates risk, outlined in \ref{subsec:General Stock by Investing by Sector}.

If the General Insights Questionnaire is selected, the user is presented with a questionnaire. % illustrated in Figure~\ref{fig:general-insights-quests1-fig} and Figure~\ref{fig:general-insights-quests2-fig}. 
Figure~\ref{fig:general-insights-answers1-fig} %and Figure~\ref{fig:general-insights-answers2-fig} 
depicts a completed questionnaire for a user who is young, single, and in debt.  \ourapproach sends the result of this questionnaire to a Perplexity.ai back-end which generates an analysis (please see Figure~\ref{fig:general-insights-results1-fig}). 

For this option, Perplexity is doing most of the work. Our  contribution is  to design the questionnaire to steer Perplexity to respond appropriately depending on the questionnaire answers. %(Figure~\ref{fig:general-insight-processing}) while processing the responses. Subsequently, Perplexity AI is called to generate a general analysis, as shown in 
 
To illustrate the result of this \say{steering}, Figure~\ref{fig:general-insights-answer3-fig} shows another completed questionnaire for a user who is financially well off, approximately 50 years old, and has a family. The corresponding Perplexity AI response is presented in Figure~\ref{fig:general-insights-results2-fig}. These examples highlight how the questionnaire responses shape the AI’s language and focus.

\FloatBarrier

% Figure page 1
\color{black}

% Figure page 2
\color{black}

% Figure page 5a
\color{black}
\renewcommand{\thefigure}{5a}
\begin{figure}[!htbp]
\centering
%\centering %% If there is a figure in wide page, please release command \centering
\begin{tikzpicture}[node distance=0.1cm and 0.7cm]
% ---- Styles ----
\tikzstyle{screenshot}=[
  rectangle, draw, thick,
  inner sep=0pt,           % no extra padding around images
  minimum width=0pt,       % let the image set the size
  minimum height=0pt,
  align=center
]
\tikzstyle{annotation}=[text width=5.5cm, align=left, font=\footnotesize]
% Screenshots - First row
\node (img1) [screenshot] at (0,0) {\includegraphics[width=0.41\textwidth]{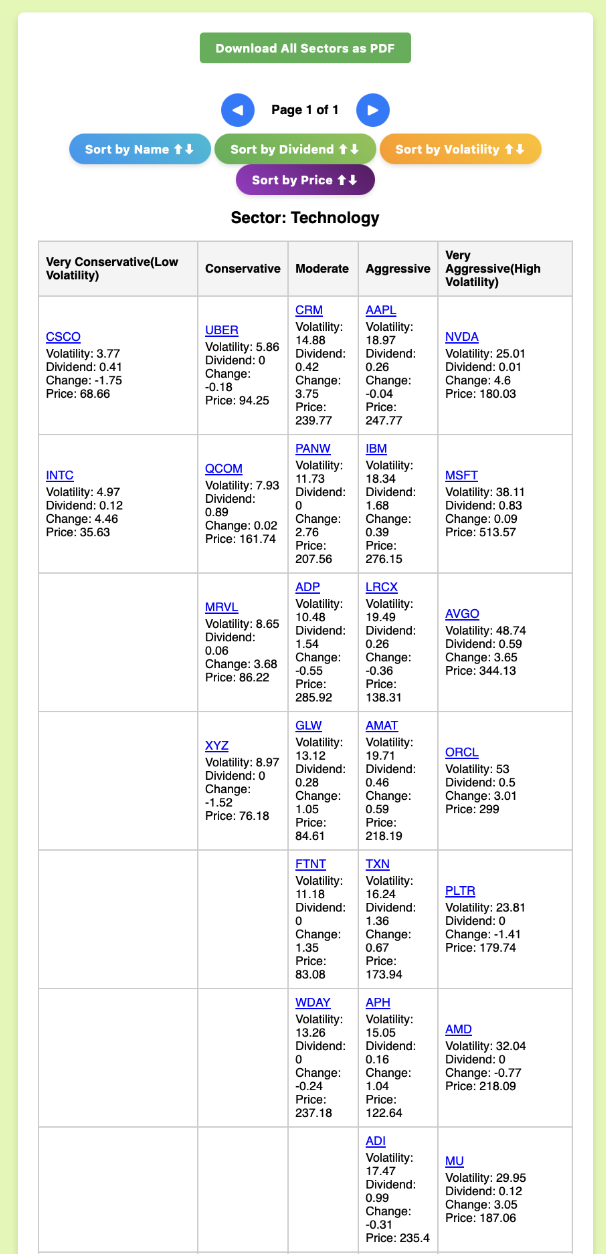}};
\node (img2) [screenshot, below=of img1] {\includegraphics[width=0.41\textwidth]{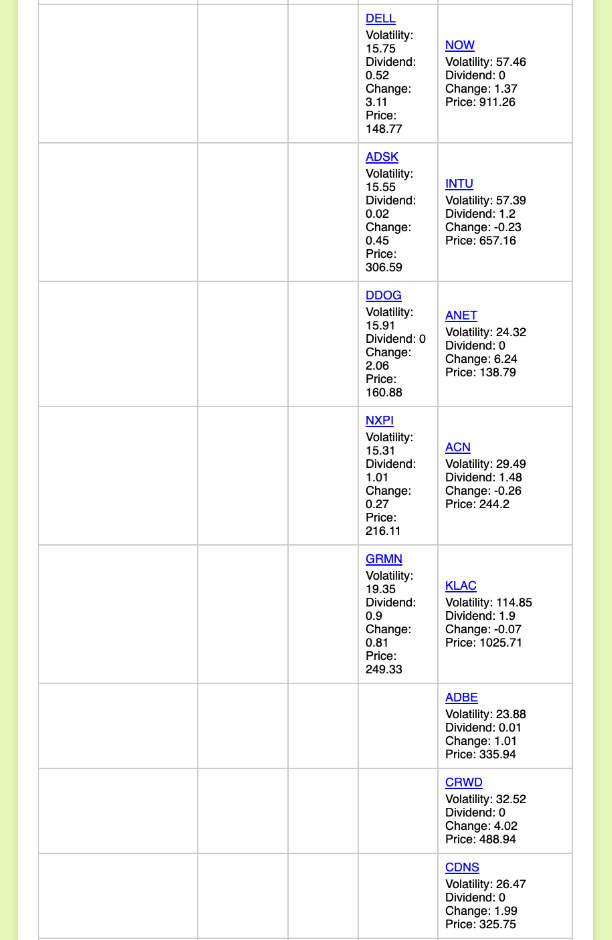}};
\end{tikzpicture}
\caption{Spreadsheet: \ourapproach generates a spreadsheet \newline containing  technology stocks.}
\label{fig:stock-results1-fig}
\end{figure}
\renewcommand{\thefigure}{\arabic{figure}}

% Figure page 5b
\color{black}
\renewcommand{\thefigure}{5b}
\begin{figure}[!htbp]
\centering
%\centering %% If there is a figure in wide page, please release command \centering
\begin{tikzpicture}[node distance=0.7cm and 0.7cm]
% ---- Styles ----
\tikzstyle{screenshot}=[
  rectangle, draw, thick,
  inner sep=0pt,           % no extra padding around images
  minimum width=0pt,       % let the image set the size
  minimum height=0pt,
  align=center
]
\tikzstyle{annotation}=[text width=5.5cm, align=left, font=\footnotesize]
% Screenshots - First row
\node (img1) [screenshot] at (0,0) {\includegraphics[width=0.90\textwidth]{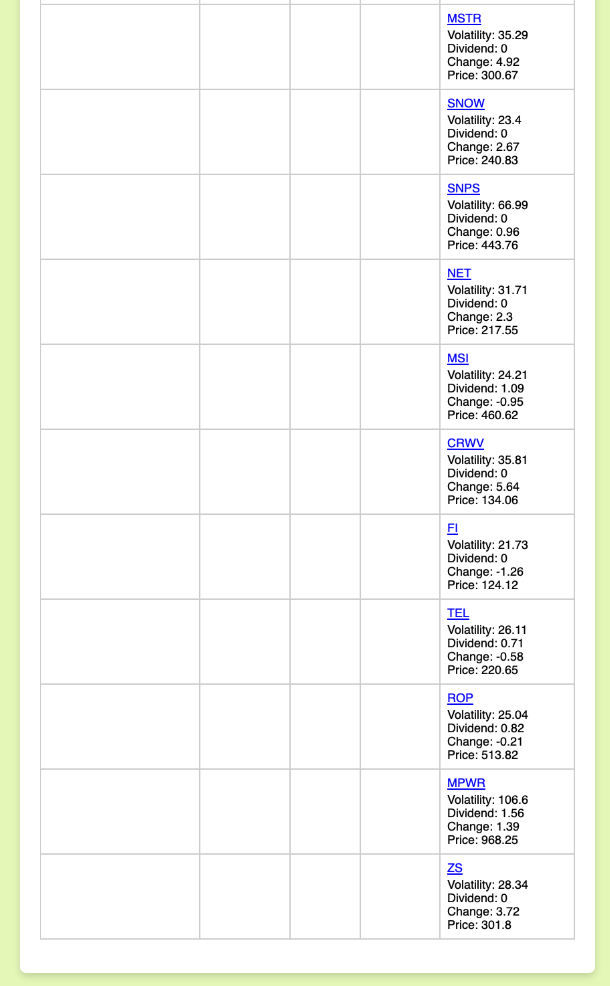}};
\end{tikzpicture}
\caption{Spreadsheet continuation containing technology stocks across all risk levels.}
\label{fig:stock-results2-fig}
\end{figure}
\renewcommand{\thefigure}{\arabic{figure}}

% Figure page 4
\color{black}

% Figure page 5
\color{black}
\renewcommand{\thefigure}{6} 
\begin{figure}[!htbp]
\centering
%\centering %% If there is a figure in wide page, please release command \centering
\begin{tikzpicture}[node distance=0.7cm and 0.7cm]
% ---- Styles ----
\tikzstyle{screenshot}=[
  rectangle, draw, thick,
  inner sep=0pt,           % no extra padding around images
  minimum width=0pt,       % let the image set the size
  minimum height=0pt,
  align=center
]
\tikzstyle{annotation}=[text width=5.5cm, align=left, font=\footnotesize]
% Screenshots - First row
\node (img1) [screenshot] at (0,0) {\includegraphics[width=0.97\textwidth]{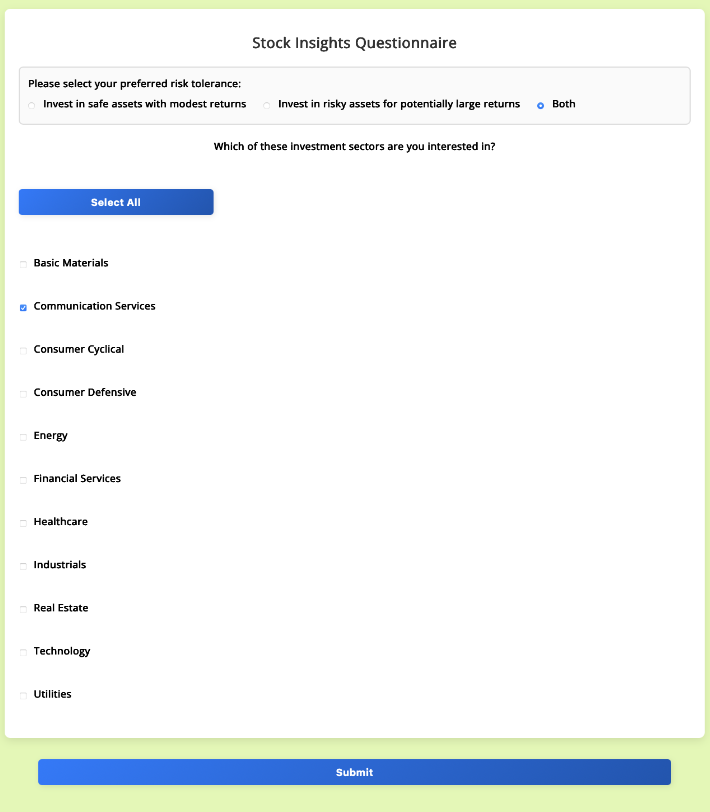}};
\end{tikzpicture}
\caption{This is a completed questionnaire for a user who is interested in communication services stocks\\ across risk levels. }
\label{fig:stock-answers2-fig}
\end{figure}

% Figure page 6a
\color{black}
\renewcommand{\thefigure}{7a} 
\begin{figure}[!htbp]
\centering
%\centering %% If there is a figure in wide page, please release command \centering
\begin{tikzpicture}[node distance=0.1cm and 0.7cm]
% ---- Styles ----
\tikzstyle{screenshot}=[
  rectangle, draw, thick,
  inner sep=0pt,           % no extra padding around images
  minimum width=0pt,       % let the image set the size
  minimum height=0pt,
  align=center
]
\tikzstyle{annotation}=[text width=5.5cm, align=left, font=\footnotesize]
% Screenshots - First row
\node (img1) [screenshot] at (0,0) {\includegraphics[width=0.45\textwidth]{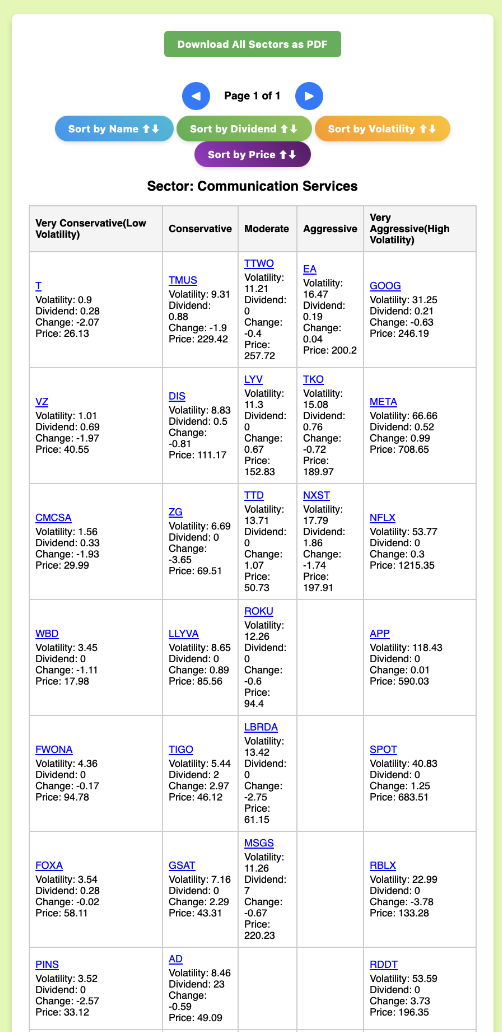}};
\node (img2) [screenshot, below=of img1] {\includegraphics[width=0.45\textwidth]{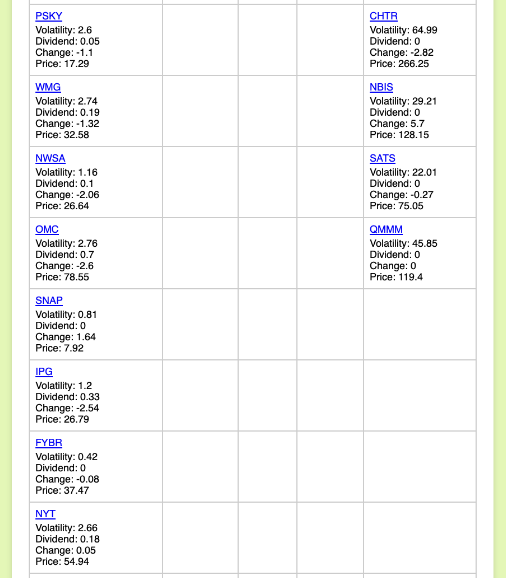}};
\end{tikzpicture}
\caption{Spreadsheet: \ourapproach generates a spreadsheet \newline containing communication stocks at all levels of risk.}
\label{fig:stock-results3-fig}
\end{figure}
\renewcommand{\thefigure}{\arabic{figure}}

% Figure page 6b
\color{black}
\renewcommand{\thefigure}{7b} 
\begin{figure}[!htbp]

\centering
%\centering %% If there is a figure in wide page, please release command \centering
\begin{tikzpicture}[node distance=0.7cm and 0.7cm]
% ---- Styles ----
\tikzstyle{screenshot}=[
  rectangle, draw, thick,
  inner sep=0pt,           % no extra padding around images
  minimum width=0pt,       % let the image set the size
  minimum height=0pt,
  align=center
]
\tikzstyle{annotation}=[text width=5.5cm, align=left, font=\footnotesize]
% Screenshots - First row
\node (img1) [screenshot] at (0,0) {\includegraphics[width=0.97\textwidth]{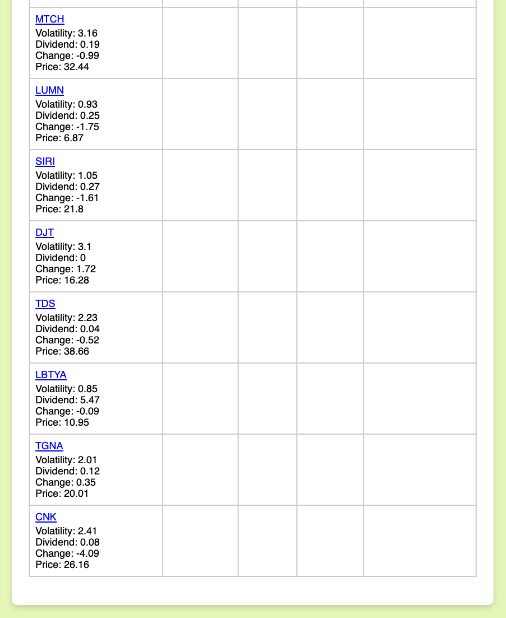}};
\end{tikzpicture}
\caption{Continuation of spreadsheet  containing communication stocks\\ over all risk categories.}
\label{fig:stock-results4-fig}
\end{figure}
\renewcommand{\thefigure}{\arabic{figure}}

% Figure page 7
\color{black}
\renewcommand{\thefigure}{8} 
\begin{figure}[!htbp]
\centering
%\centering %% If there is a figure in wide page, please release command \centering
\begin{tikzpicture}[node distance=0.7cm and 0.7cm]
% ---- Styles ----
\tikzstyle{screenshot}=[
  rectangle, draw, thick,
  inner sep=0pt,           % no extra padding around images
  minimum width=0pt,       % let the image set the size
  minimum height=0pt,
  align=center
]
\tikzstyle{annotation}=[text width=5.5cm, align=left, font=\footnotesize]
% Screenshots - First row
\node (img1) [screenshot] at (0,0) {\includegraphics[width=0.87\textwidth]{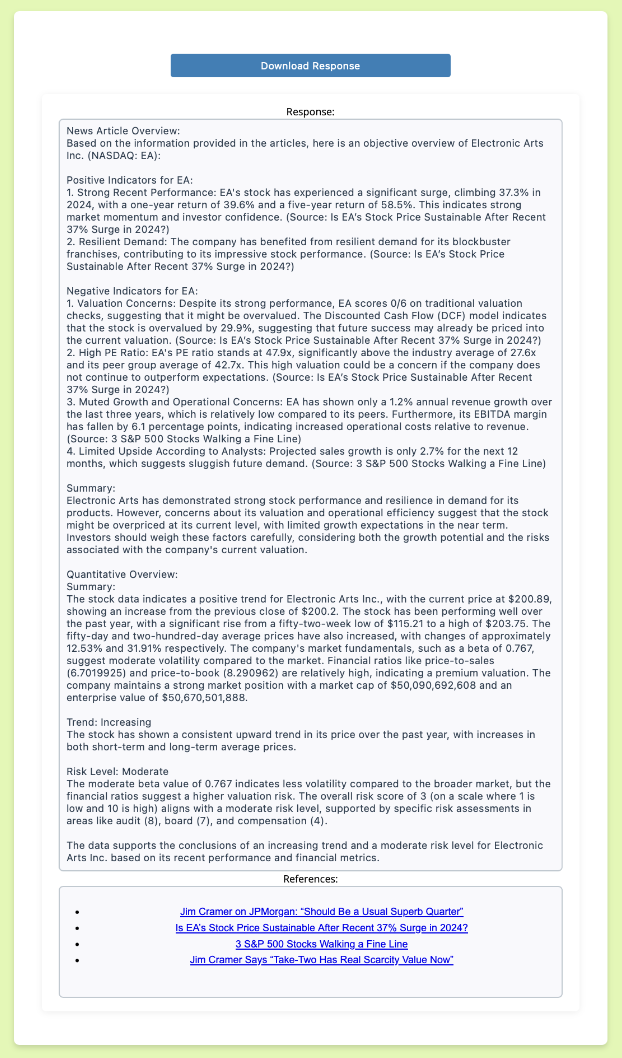}};
\end{tikzpicture}
\caption{Stock insights results about Electronic Arts (EA).}
\label{fig:EA-answer-fig}
\end{figure}

% Figure page 8
\color{black}

% Figure page 9a
\color{black}
\renewcommand{\thefigure}{9}
\begin{figure}[!htbp]
\centering
%\centering %% If there is a figure in wide page, please release command \centering
\begin{tikzpicture}[node distance=0.1cm and 0.7cm]
% ---- Styles ----
\tikzstyle{screenshot}=[
  rectangle, draw, thick,
  inner sep=0pt,           % no extra padding around images
  minimum width=0pt,       % let the image set the size
  minimum height=0pt,
  align=center
]
\tikzstyle{annotation}=[text width=5.5cm, align=left, font=\footnotesize]
% Screenshots - First row
\node (img1) [screenshot] at (0,0) {\includegraphics[width=0.48\textwidth]{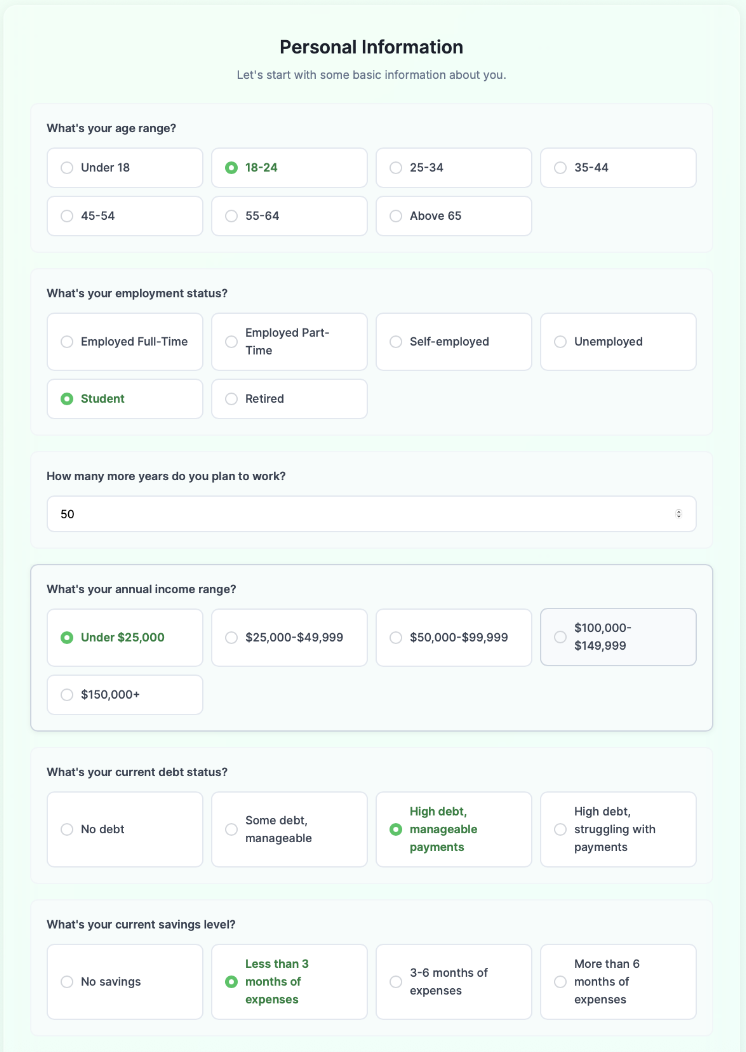}};
\node (img2) [screenshot, below=of img1] {\includegraphics[width=0.48\textwidth]{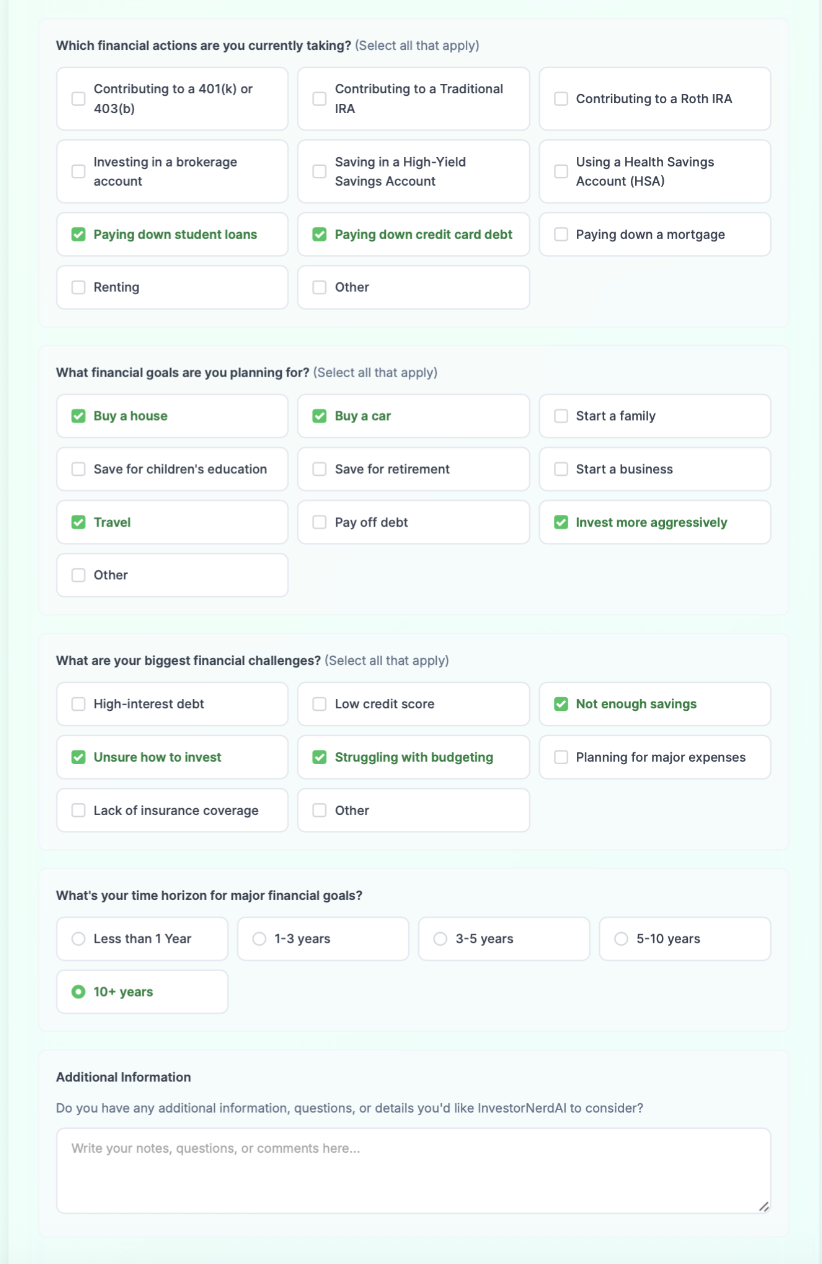}};
\end{tikzpicture}
\caption{This is a completed questionnaire for a user who is young, single, and in debt.}
\label{fig:general-insights-answers1-fig}
\end{figure}
\renewcommand{\thefigure}{\arabic{figure}}

% Figure page 9
\color{black}
\renewcommand{\thefigure}{10} 
\begin{figure}[!htbp]
\centering
%\centering %% If there is a figure in wide page, please release command \centering
\begin{tikzpicture}[node distance=0.7cm and 0.7cm]
% ---- Styles ----
\tikzstyle{screenshot}=[
  rectangle, draw, thick,
  inner sep=0pt,           % no extra padding around images
  minimum width=0pt,       % let the image set the size
  minimum height=0pt,
  align=center
]
\tikzstyle{annotation}=[text width=5.5cm, align=left, font=\footnotesize]
% Screenshots - First row
\node (img1) [screenshot] at (0,0) {\includegraphics[width=0.75\textwidth]{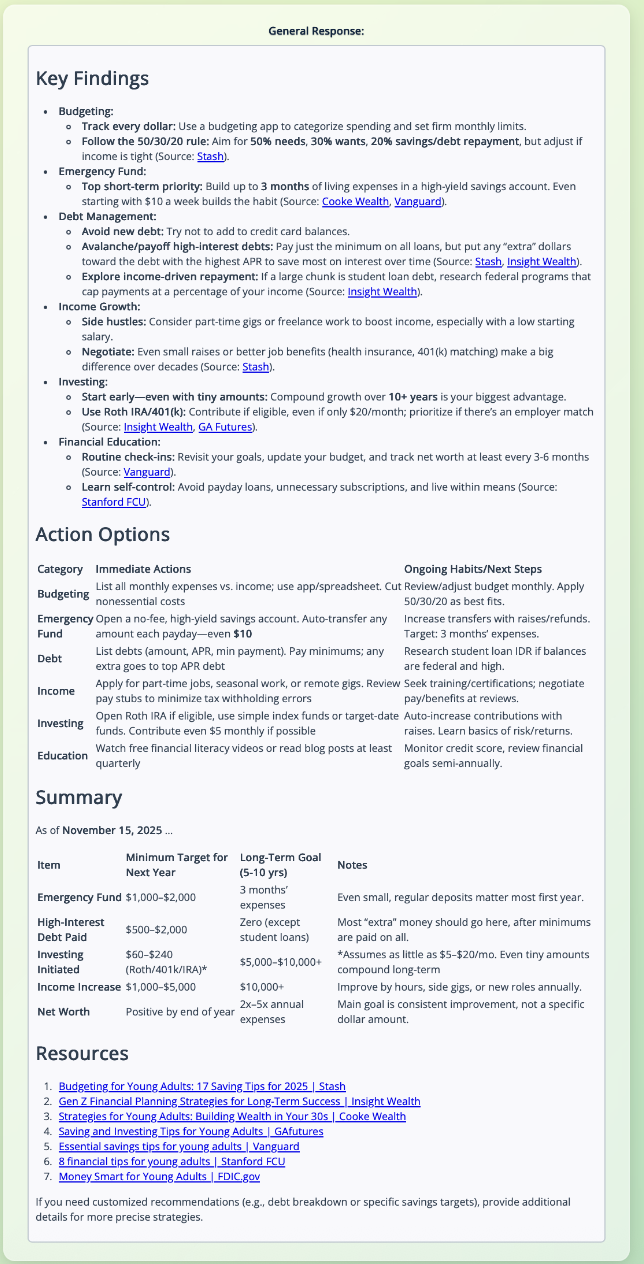}};
\end{tikzpicture}
\caption{General insights results for a young, single and in debt person.}
\label{fig:general-insights-results1-fig}
\end{figure}

% Figure page 11
\color{black}
\renewcommand{\thefigure}{11} 
\begin{figure}[!htbp]
\centering
%\centering %% If there is a figure in wide page, please release command \centering
\begin{tikzpicture}[node distance=0.1cm and 0.7cm]
% ---- Styles ----
\tikzstyle{screenshot}=[
  rectangle, draw, thick,
  inner sep=0pt,           % no extra padding around images
  minimum width=0pt,       % let the image set the size
  minimum height=0pt,
  align=center
]
\tikzstyle{annotation}=[text width=5.5cm, align=left, font=\footnotesize]
% Screenshots - First row
\node (img1) [screenshot] at (0,0) {\includegraphics[width=0.48\textwidth]{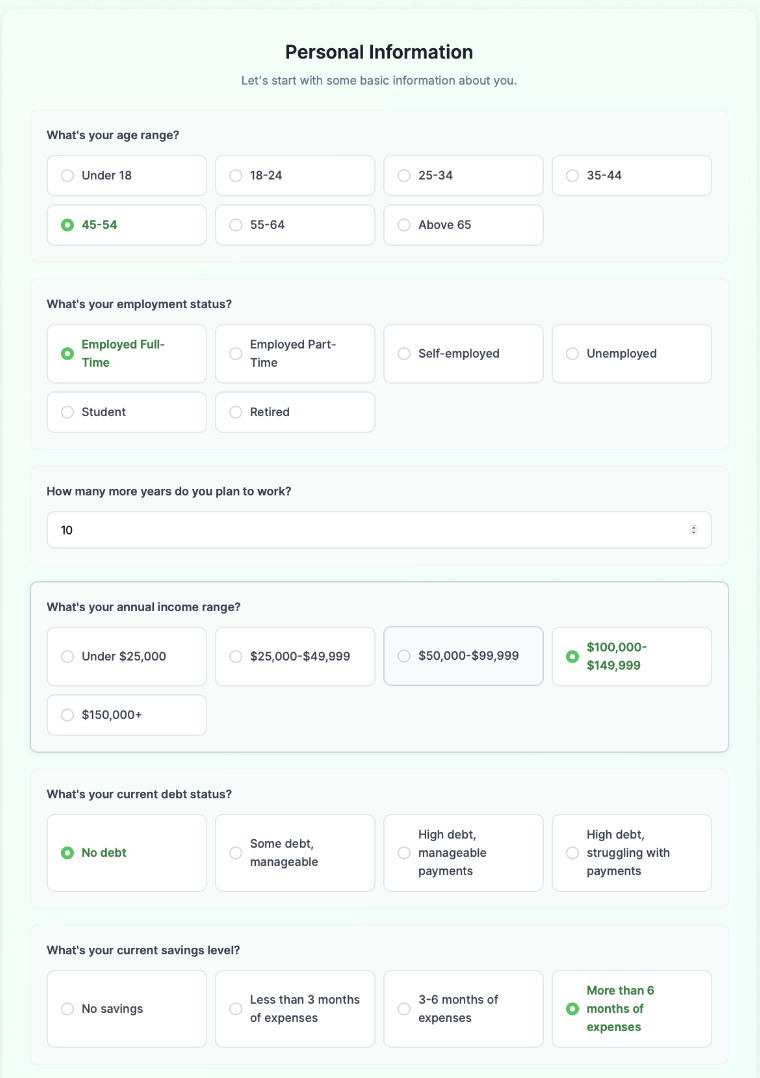}};
\node (img2) [screenshot, below=of img1] {\includegraphics[width=0.48\textwidth]{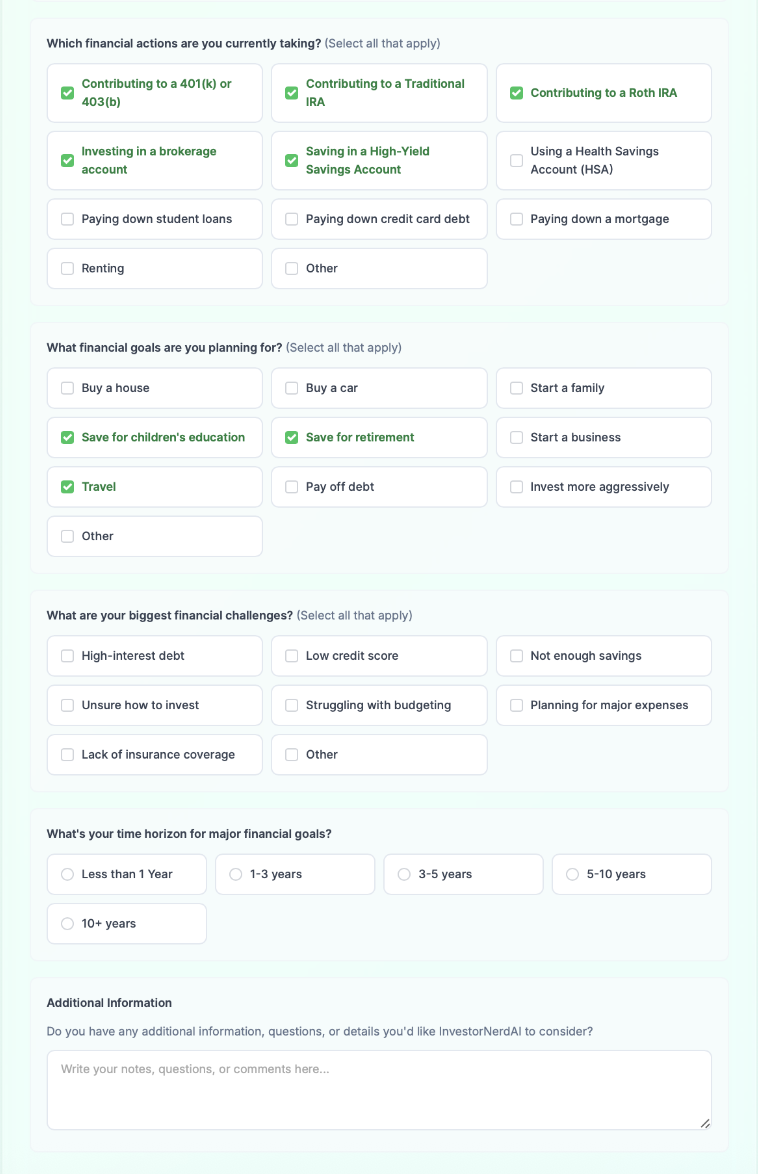}};
\end{tikzpicture}
\caption{This is a completed questionnaire for a user who is financially well off, approximately 50 years old, and \newline has a family.}
\label{fig:general-insights-answer3-fig}
\end{figure}

% Figure page 12
\color{black}
\renewcommand{\thefigure}{12} 
\begin{figure}[!htbp]
\centering
%\centering %% If there is a figure in wide page, please release command \centering
\begin{tikzpicture}[node distance=0.7cm and 0.7cm]
% ---- Styles ----
\tikzstyle{screenshot}=[
  rectangle, draw, thick,
  inner sep=0pt,           % no extra padding around images
  minimum width=0pt,       % let the image set the size
  minimum height=0pt,
  align=center
]
\tikzstyle{annotation}=[text width=5.5cm, align=left, font=\footnotesize]
% Screenshots - First row
\node (img1) [screenshot] at (0,0) {\includegraphics[width=0.70\textwidth]{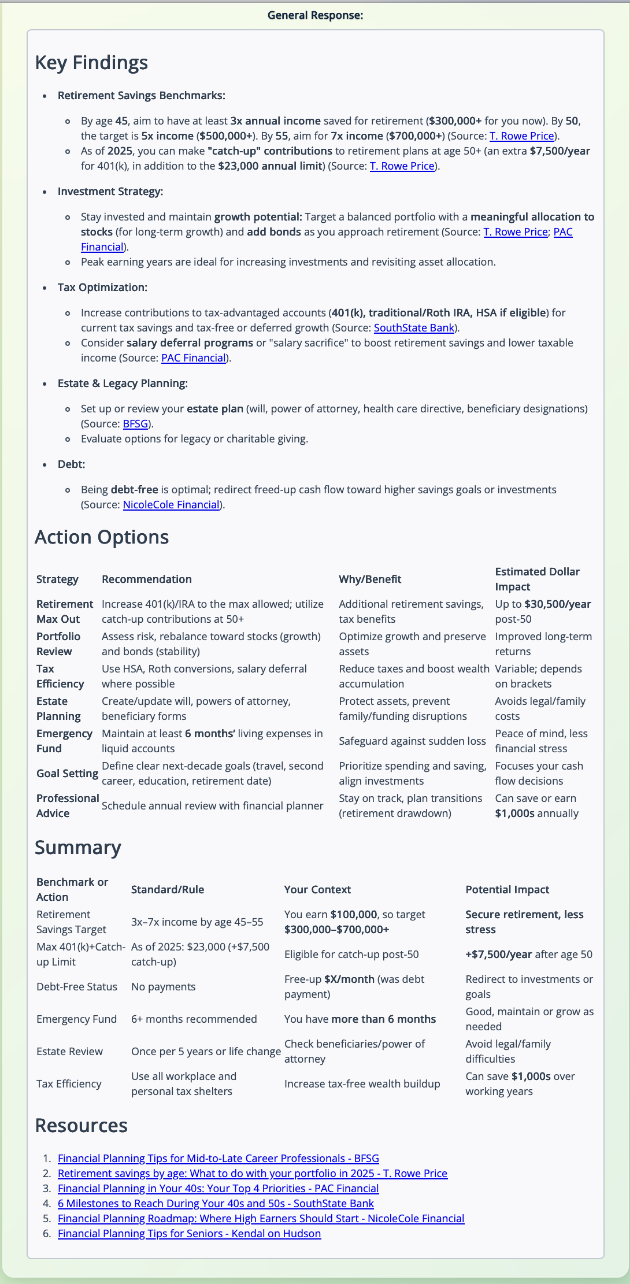}};
\end{tikzpicture}
\caption{General insights results for the financially secure 50 year old.}
\label{fig:general-insights-results2-fig}
\end{figure}

% Figure page 16
\color{black}

\clearpage

\section{Related Work}
\label{sec:related-work}

The following are the highest quality and accessible investment information systems that are open to the public or have a free tier that we have found. As of mid-2025, the systems that are closest to \ourapproach in Functionality are AI Financial Planning, Kavout, You.com, and FinanceWizard.

\paragraph{Kavout}
Kavout is an AI-driven investment analytics platform that applies machine learning models to market and fundamental data in order to generate predictive stock rankings, most notably through its proprietary Kai Score. The system emphasizes quantitative pattern recognition and factor-based modeling to forecast stock performance. However, Kavout primarily presents outputs as numerical scores or rankings, offering limited explanation of contributing factors. The platform does not provide conversational interaction, user-specific synthesis, or explicit risk categorization, requiring users to independently interpret model outputs.

\paragraph{You.com}
You.com is a general-purpose AI-powered search and question-answering engine that supports conversational queries across a wide range of domains, including finance. For financial queries, the system retrieves and summarizes publicly available information from multiple sources in a natural language format. While this conversational interface improves accessibility, You.com is not specialized for investment analysis. It does not incorporate investor profiling, structured risk assessment, or domain-specific financial modeling, and its responses are not tailored to long-term investment decision-making.

\paragraph{FinanceWizard}
FinanceWizard represents a class of consumer-focused AI tools designed to simplify financial concepts through conversational explanations and high-level guidance. The platform emphasizes ease of use and accessibility, often providing brief summaries or recommendations in response to user questions. However, FinanceWizard does not support deep personalization, formal risk modeling, or systematic comparison across risk categories. Its outputs are primarily descriptive rather than analytical, limiting its applicability for investors seeking structured and repeatable decision support.

\paragraph{AI Financial Planning}
AI Financial Planning is a GPT-powered conversational tool that addresses general personal finance questions and introductory investment concepts. The platform pairs a natural language interface with a curated directory of third-party financial products, directing users toward external services in response to their queries. While the system supports open-ended dialogue, it 
does not perform stock analysis, risk profiling, or portfolio modeling. Its outputs consist of general educational responses and product referrals rather than structured investment guidance, and the platform explicitly disclaims the provision of personalized financial 
or investment advice.

\section{Materials and Methods}\label{materials-methods}

We first state the principles underlying our design and then explain the implementation.

\subsection{Design Principles of \ourapproach}

 %In order to enhance  LLM performance and adaptability, a few tactics are available. 
 \ourapproach makes extensive use of two concepts to support accuracy and adaptability in financial contexts: \textbf{structured contextual prompting} and \textbf{integration of external retrieval mechanisms}. 
 
 Structured contextual prompting involves crafting prompts with predefined templates that  organize user inputs into fields that the model can interpret unambiguously.  In the absence of this technique, a user might type, for example: \texttt{"Can I invest in tech stocks? I’m 25 and not afraid of risk."}.  The model would then have to infer context from unstructured prose, risking misinterpretation. With structured contextual prompting, the same request is transformed into an  input format such as:
\begin{verbatim}
Age: 25
Employment status: Full-time
Risk tolerance: High
Investment focus: Technology sector
Question: Can I invest in tech stocks?
\end{verbatim}
This structured form ensures that the model receives explicit, machine-readable parameters, enabling  tailored responses. 

\ourapproach applies this idea by encoding all user inputs (e.g., risk tolerance, investment horizon, and financial constraints) into predefined prompt templates that guide the LLM toward generating structured and relevant responses. These templates explicitly specify financial goals, constraints, and risk parameters. Full prompt templates used by the system are provided in Appendix~\ref{prompt-appendix}.

External retrieval mechanisms pertain to a pipeline that dynamically queries and integrates external data sources into the model’s context at inference time. In our case, the primary source is Yahoo Finance (YFinance), which provides both real-time market data and access to relevant historical datasets. Retrieved information is parsed, summarized, and then queried by   the prompt so the LLM can ground its answers in both up-to-date market conditions and a broader financial context. %Specifically, the real-time financial data comes from APIs (e.g., Yahoo Finance) and recent news articles. % to augment the LLM's knowledge base, ensuring factual and current outputs.  %Unlike generic real-time sources, these mechanisms are explicitly linked to the model’s reasoning process—retrieved information is parsed, summarized, and embedded into the prompt so the LLM can ground its answers in both current and contextually relevant domain knowledge.

\ourapproach adopts %based on layered prompt design and modular integration. Instead of a rigid generation pipeline, \ourapproach system is 
a structured approach divided into discrete stages: user input collection, financial data retrieval, contextual prompt assembly, and response generation. This form of prompt  chaining between modules helps achieve coherence in system output. %To maintain consistency and information integrity, there are limits on speculative language and requirements for citations. 
In addition, \ourapproach's responses are framed as educational summaries rather than advice. %, and the system will explicitly state uncertainty when underlying data or information is incomplete or ambiguous. \textcolor{red}{John NOTES: This quote "the system will explicitly state uncertainty when underlying data or information is incomplete or ambiguous. " is not true as the point of the questionnaire is meant to provide enough context to the AI, also we don't have it return an error message to include more info.}
These methods allow \ourapproach (i) to personalize its responses to specific user profiles, (ii) to adapt its response to evolving financial news, and (iii) to make use of  authoritative  data in a coherent way.

%This modular approach supports a safety  framework that does not rely on model retraining or human moderation. This structure provides interpretability and control, particularly when coupled with clearly defined financial thresholds and transparency protocols. 

%In summary, \ourapproach combines dynamic prompt generation, real-time data integration, and multi-stage output processing to create a financial information system that aims to educate   non-expert users. %The platform is openly accessible and emphasizes interpretability and education over automation or prediction, distinguishing it from other commercial tools that prioritize trading optimization or portfolio performance.

%\textcolor{cyan}{We might point out that another area of relevant work is RAG (retrieval augmented generalization) which entails going to primary sources.} 

 %Although this approach does not follow the traditional Retrieval-Augmented Generation (RAG) framework, it achieves a similar functionality by enabling the integration of external data sources and up-to-date financial information into the generation process.

\subsection{Implementation}

\ourapproach is built on FastAPI and integrates real-time financial data from Yahoo Finance (yfinance), large language models (ChatGPT or Perplexity), and interactive questionnaire logic to generate financial insights for users. \ourapproach supports three primary workflows, based on the three choices from the landing page: (1) the Stock Insights for individual stock analysis, (2) the Stock Insights Questionnaire for sector-based investment classification, and (3) the General Insights Questionnaire for general insight generation.  For (1) and (2) we use ChatGPT and for (3) we use Perplexity.  
%\textcolor{blue}{FIXED} 
%\textcolor{blue}{FIXED} \textcolor{red}{%John NOTES: When it comes to LLMs, we use either ChatGPT or Perplexity given the specific tool used, also 2), the choice name is not correct, currently it is called Stock questionnaire, but this needs to be changed on the website if we want to rename in that way. Dennis agrees that 
%Be sure we call each choice whatever it is called on the website.  }

All workflows deliver real-time feedback   to the user. The responses contain financial summaries, risk classifications, and article links. These responses are parsed from back-end JSON responses and provided as interactive tables or expandable markdown sections. %The interface changes based on what the user selects, making it easy to collect information and guide them through the right steps, while staying perfectly synced with the server.

%The landing page of frontend of \ourapproach presents users with two options: \textcolor{red}{Why "two"?} \texttt{"General Insight"} and \texttt{"Stock Insight"}.

\subsection{General Stock Investing by Sector} \label{subsec:General Stock by Investing by Sector}

\begin{figure}[H]
\centering
\begin{tikzpicture}[node distance=1.5cm, every node/.style={align=center}]
\node (start) [draw, rounded corners, fill=blue!15] {Landing Page};
\node (choice) [below of=start, draw, rounded corners, fill=green!15] {User selects Stock Insights Questionnaire};
\node (inputs) [below of=choice, draw, rounded corners] {Select sectors \& risk tolerance};
\node (compute) [below of=inputs, draw, rounded corners] {Map to risk bands + yfinance lookup};
\node (classify) [below of=compute, draw, rounded corners, fill=orange!20] {Classify top stocks by volatility};
\node (result) [below of=classify, draw, rounded corners, fill=yellow!20] {Display structured results};
\draw[->] (start) -- (choice);
\draw[->] (choice) -- (inputs);
\draw[->] (inputs) -- (compute);
\draw[->] (compute) -- (classify);
\draw[->] (classify) -- (result);
\end{tikzpicture}
\caption{Workflow for Sector-Based Stock Classification}
\label{workflow-fig_sector-based}
\end{figure}
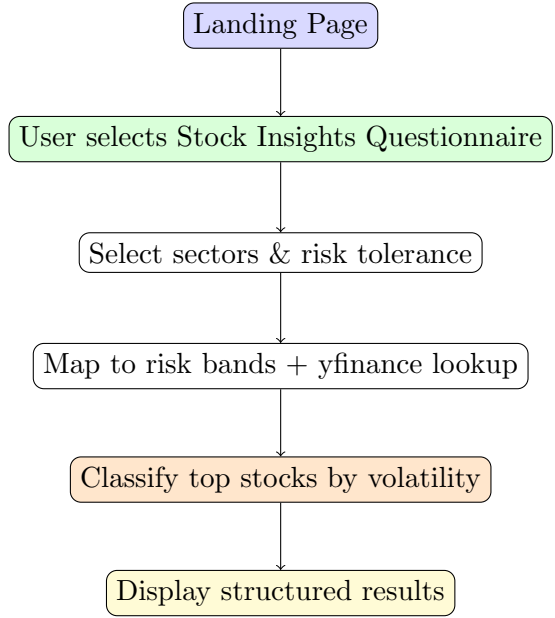

In \texttt{"Stock Insight Questionnaire"} mode, \ourapproach prompts the user to select one or more market sectors (e.g., financials, technology) as well as a risk tolerance level. The front end captures these inputs and submits them to the back end, then the back end invokes a risk classifying function to separate each selected sector's stocks into risk categories. The information comes from \texttt{yfinance}. This is all shown in Algorithm \ref{sector-based-alg} and Figure \ref{workflow-fig_sector-based}.  Each stock’s risk is computed by taking standard deviation of its daily closing price over roughly the past six months of trading data. Then the stocks are assigned to five risk categories using fixed volatility thresholds: \(\leq\) 5 (Very Conservative), 5–10 (Conservative), 10–15 (Moderate), 15–20 (Aggressive), and $>$20 (Very Aggressive). As shown in the Appendix \ref{accuracy-analysis} the riskiness is 'sticky' in the sense that risky stocks tend to stay risky and conservative stocks remain conservative. Although not surprising, the numerical results are still interesting.

After \ourapproach lists the stocks that satisfy the user's request in the form of a spreadsheet. The user may then choose specific stocks to get an analysis based on real-time data and analysis of the stock as described in Section \ref{single-stock-section} whose workflow is shown in Figure \ref{workflow-fig_sector-based} and whose user interface is shown in Figure~\ref{fig:stock-results1-fig}, Figure~\ref{fig:stock-results2-fig}, Figure~\ref{fig:stock-results3-fig}, and Figure~\ref{fig:stock-results4-fig}.

\begin{algorithm}[H]
    \caption{Sector Classification Workflow}
    \label{sector-based-alg}
    \begin{algorithmic}[1]
        \State User selects ``Stock Insights Questionnaire'' option
        \State Prompt user for risk tolerance and preferred sectors (e.g., Technology, Financial)
        \State Determine allowed risk classes based on tolerance mapping
        \For{\textbf{each} sector \textbf{in} selected sectors}
            \State Fetch top 50 companies in sector using \texttt{yfinance}
            \State Compute 6-month volatility, dividend yield, and price change per ticker
            \State Classify each ticker into risk bands (Very Conservative to Very Aggressive)
        \EndFor
        \State Filter tickers to those within allowed risk classes
        \State Sort results from least to most volatility; within each volatility band, rank by 6-month price increase
        \State Display structured JSON output to user
    \end{algorithmic}
\end{algorithm}

\subsection{Algorithm 3: Single Stock Analysis via GPT and News}\label{single-stock-section}
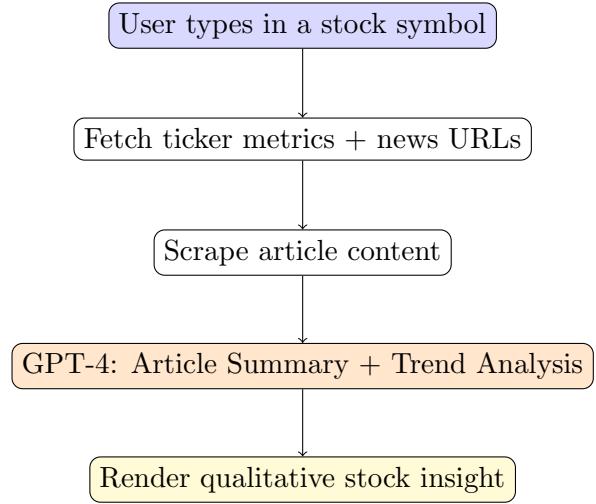
\begin{figure}[H]
\centering
\begin{tikzpicture}[node distance=1.5cm, every node/.style={align=center}]
\node (click) [draw, rounded corners, fill=blue!15] {User types in a stock symbol};
\node (fetch) [below of=click, draw, rounded corners] {Fetch ticker metrics + news URLs};
\node (scrape) [below of=fetch, draw, rounded corners] {Scrape article content};
\node (synth) [below of=scrape, draw, rounded corners, fill=orange!20] {GPT-4: Article Summary + Trend Analysis};
\node (display) [below of=synth, draw, rounded corners, fill=yellow!20] {Render qualitative stock insight};
\draw[->] (click) -- (fetch);
\draw[->] (fetch) -- (scrape);
\draw[->] (scrape) -- (synth);
\draw[->] (synth) -- (display);
\end{tikzpicture}
\caption{Workflow for Single Stock Analysis}
\label{single-stock-fig}
\end{figure}

\textbf{Objective:} When a user clicks a specific stock, either based on the sector workflow or by directly querying the landing page, \ourapproach conducts a qualitative analysis via scraped news and yfinance metrics using GPT-4 Turbo. The algorithm is shown in  Algorithm \ref{single-stock-alg} whose workflow is shown in Figure \ref{single-stock-fig} and whose user interface is shown in Figure~\ref{fig:Nvidia-answer-fig} and Figure~\ref{fig:EA-answer-fig}. 

\begin{algorithm}[H]
\caption{Single Stock Drill-Down Workflow}
\begin{algorithmic}[1]
\State User types in a stock symbol 
\State Fetch ticker fundamentals from yfinance (price, EPS, margin, etc.)
\State Fetch top 5 news URLs from yfinance’s news module
\State Scrape and truncate content via BeautifulSoup
\State Synthesize news summary with \texttt{gpt-4-turbo}
\State Analyze ticker metrics with \texttt{gpt-4-turbo} for trend/risk
\State Combine insights and display full stock analysis
\end{algorithmic}
\label{single-stock-alg}
\end{algorithm}

%\textcolor{blue}{fIXED}\textcolor{red}{John Notes: The user does not click on a stock, it queues it up or types it out, when it comes to the State 1 on Figure 4}
\subsection{Algorithm: General Insight Workflow via Questionnaire}
%\textcolor{blue}{FIXED} \textcolor{red}{This should be after the stock insights. Please change the order of the figures accordingly.}
This choice is aimed at collecting   data about a hypothetical individual's financial profile through a structured form and generating holistic investment insight  using Perplexity AI \url{perplexity.ai} 

\begin{figure}[H]
\centering
\begin{tikzpicture}[node distance=1.5cm, every node/.style={align=center}]
\node (start) [draw, rounded corners, fill=blue!15] {Landing Page};
\node (form) [below of=start, draw, rounded corners, fill=green!15] {User selects General Insight};
\node (questionnaire) [below of=form, draw, rounded corners] {9-question form \& optional general query};
\node (prompt) [below of=questionnaire, draw, rounded corners] {Prompt construction};
\node (perplexity) [below of=prompt, draw, rounded corners, fill=orange!20] {Query Perplexity API};
\node (result) [below of=perplexity, draw, rounded corners, fill=yellow!20] {Display:\newline Key Findings, Actions, Sources};
\draw[->] (start) -- (form);
\draw[->] (form) -- (questionnaire);
\draw[->] (questionnaire) -- (prompt);
\draw[->] (prompt) -- (perplexity);
\draw[->] (perplexity) -- (result);
\end{tikzpicture}
\caption{Workflow for General Insight via Questionnaire and Perplexity API}
\label{general-insight-fig}
\end{figure}
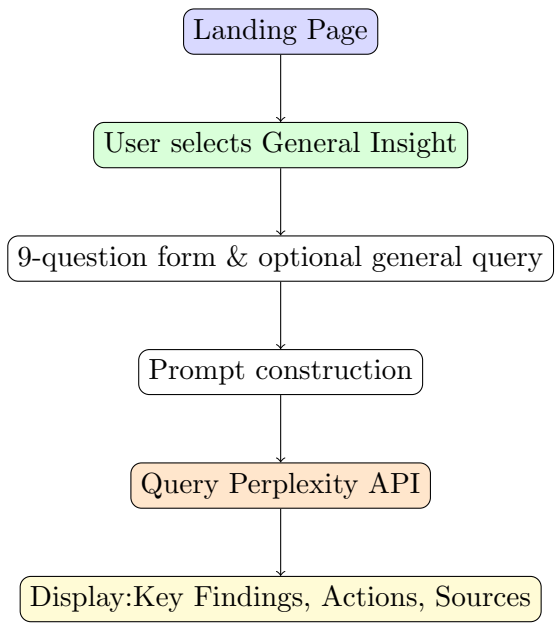

The front end presents the users with a questionnaire composed of %multiple input modalities, including checkbox groups and text fields. These inputs mirror the \texttt{FormData} schema in the FastAPI back end. Questions include 
questions about a hypothetical person's  age group, income, savings, debt status, planning goals, and risk tolerance. As seen earlier, Figure \ref{fig:general-insights-answers1-fig} %and \ref{fig:general-insights-answers2-fig} 
shows one complete example for a young person having few assets. %\textcolor{blue}{because there are too many of them I'll just add a reference to Investornerd in action, is this ok?}\textcolor{red}{Create a box with the questions and choices and refer to that} 
Input values are captured and submitted in JSON format to the back end's \texttt{/questionnaire} endpoint. Then the back end routes the information to the Perplexity API for analysis as shown in Algorithm \ref{general-insight-alg} whose workflow is shown in Figure \ref{general-insight-fig} and whose result for that same young person is shown in Figure \ref{fig:general-insights-results1-fig}. %\textcolor{red}{John NOTES: We need to link the inteface figure here to the general insights screenshot I think}

\begin{algorithm}[H]
\caption{General Insight Workflow}
\begin{algorithmic}[1]
\State User selects the ``General Insights Questionnaire'' option
 \State Display and collect responses to a financial questionnaire
\State Construct a natural language prompt including all answers
\State Append user’s custom general financial question
\State Query \texttt{sonar-pro} model from Perplexity API with the full prompt
\State Receive structured financial insight with sources
\State Display action options, and links
\end{algorithmic}
\label{general-insight-alg}
\end{algorithm}

\section{\ourapproach Compared to Other State-of-the-Art Systems}

The assessment of \ourapproach compared to the state of the art comprises two components:

\begin{enumerate}
\item This section presents a multi-dimensional functionality comparison of \ourapproach with other state-of-the-art systems. The new functionality is our primary contribution.
 %   \item Section \ref{accuracy-analysis} describes a data-driven evaluation of how well \ourapproach's risk classification (based on volatility) as of time period $P$ aligns with those same stocks   at time period $P+1$. 
    % \item \textcolor{red}{Dennis thinks we can add a table where we compare the functionality of our tools a whole including the stock analysis with other financial planning tools. I think we can say that we are a better one-stop shop.}

 \item Regarding the functionality that \ourapproach shares with other financial education systems, namely general financial insight, two sections compare our system with the state of the art. Section \ref{sts-testing} provides a Semantic Textual Similarity (STS) evaluation \cite{reimers2019} of how closely \ourapproach recommendations align with the responses generated by other AI financial planning systems.   %experiment involving a single domain expert as well as lay users and other state-of-the-art systems. The sample size is small and is meant only to give an anecdotal comparison.
    % \textcolor{red}{John Notes: **MAKE SURE WE ADD THE EXPERT DATA HERE** Dennis note: Just give a qualitative comparison}
  
    % \textcolor{red}{Dennis is wondering whether this part makes sense}
    % \item Section \ref{hallucination} describes the results of a manual hallucination check to see whether assertions in the summaries of state-of-the-art systems as well as \ourapproach that cite a reference accurately characterize  that reference.
\end{enumerate}

%These analyses allow us to assess \ourapproach's performance from both a quantitative and qualitative perspective. %\textcolor{cyan}{Shela: in each of the following sections, we should describe the setup and the results for each of the three tests} %The double-blind setup for the manual evaluation minimizes potential biases, while the involvement of registered dietitians ensures that the assessment is grounded in practical, domain-specific considerations.

\subsection{Functionality Comparison of \ourapproach with the State of the Art}

% The primary contribution of \ourapproach compared to the state of the art lies in the functionality we offer users.

State-of-the-art AI financial planning and advisory platforms focus on high-level, goal-driven financial guidance such as retirement planning, budgeting, and asset allocation. %These systems rely on structured questionnaires to infer investor preferences and risk tolerance, producing portfolio recommendations optimized for long-term objectives. 
While effective for broad financial planning, they  do not support individual stock-level exploration, qualitative synthesis of financial news, or explicit breakdowns of risk across multiple dimensions. %Interpretability is often limited, as recommendations are generated through predefined optimization frameworks rather than interactive analysis. 

\ourapproach  tries to make information about particular financial assets interpretable and includes a risk assessment based on historical data. Our approach does not  provide a chat feature. We avoid keeping state about users to protect their privacy.  Table \ref{systemcomparison} summarizes the functionality comparison between \ourapproach and other state-of-the-art systems. Our main contribution compared to other financial education system is introducing a simple document-based analysis of both individual stocks and sectors of stocks.

\begin{table}[H]
    \caption{\textbf{System Functionality Comparison.}
    \ourapproach includes individual stock analysis and sector-based risk analyses absent from other state-of-the-art systems.
    Some other systems include a conversational chat feature which we have avoided for privacy reasons.}
    \centering
    \small
    \setlength{\tabcolsep}{4pt}
    \resizebox{\textwidth}{!}{%
    \begin{tabular}{lccccc}
    \toprule
    \textbf{System}
    & \textbf{\shortstack{Quantitative\\Stock\\Analysis}}
    & \textbf{\shortstack{Stock\\Analysis\\by Sector}}
    & \textbf{\shortstack{Stock\\Analysis\\by Risk}}
    & \textbf{Conversational}
    & \textbf{\shortstack{General\\Financial\\Education}} \\
    \midrule
    AI Financial Planning  &   &   &   &   & \checkmark \\
    Kavout                 &   &   &   &   & \checkmark \\
    You.com                &   &   &   & \checkmark & \checkmark \\
    FinanceWizard          &   &   &   & \checkmark & \checkmark \\
    \ourapproach           & \checkmark & \checkmark & \checkmark &   & \checkmark \\
    \bottomrule
    \end{tabular}
    }
    \label{systemcomparison}
\end{table}

\subsection{Stock Classification Tools}

To  contextualize the stock classification feature of \ourapproach, Table \ref{tab:sota-comparison} presents a targeted comparison against state-of-the-art AI-driven investment platforms that perform some form of stock scoring or categorization. Unlike the broader functionality comparison in Table \ref{systemcomparison}, this comparison focuses specifically on whether existing systems support explicit risk-tier classification, investor profiling, and explainable outputs --- the dimensions most relevant to our contribution. As the table shows, existing platforms that do classify 
stocks rely on numerical scores or buy/sell signals rather than named risk categories, and none combine explicit risk classification with investor profiling and explainable outputs in a single system.

\begin{table*}[ht]
\centering
\renewcommand{\arraystretch}{1.35}
\caption{\textbf{Stock Classification Comparison.}
Comparison of state-of-the-art AI-driven investment platforms across dimensions 
relevant to structured stock classification and investor-aligned risk categorization.}
\label{tab:sota-comparison}
\resizebox{\textwidth}{!}{%
\begin{tabular}{lccccc}
\toprule
\textbf{System} &
\textbf{\shortstack{Stock\\Classification}} &
\textbf{\shortstack{Explicit Risk\\Categories}} &
\textbf{\shortstack{Investor\\Profiling}} &
\textbf{\shortstack{Fundamental\\Analysis}} &
\textbf{\shortstack{Free\\Tier}} \\
\midrule
Kavout             & \checkmark ~(Kai Score 1--9)             & \checkmark &  & \checkmark &  \\
Danelfin          & \checkmark ~(AI Score 1--10)             &  &  & \checkmark & \checkmark \\
Seeking Alpha & \checkmark~(Strong Buy -- Strong Sell)  &  &  & \checkmark & \checkmark \\
Finviz           & \checkmark~(sector/fundamental filters) &  &  & \checkmark & \checkmark \\
Yahoo Finance Screener & \checkmark~(sector/technical filters) &  &  & \checkmark & \checkmark \\
\ourapproach & \shortstack{\checkmark~(conservative/\\moderate/aggressive)} & \checkmark & \checkmark & \checkmark & \checkmark \\
\bottomrule
\end{tabular}%
}

\smallskip
\begin{flushleft}
\footnotesize
\textit{Stock Classification}: Whether the platform assigns stocks to categories or scores, and the type of classification used.

\textit{Explicit Risk Categories}: Whether outputs are labeled using named investor risk 
tiers (e.g., conservative, aggressive) rather than numerical scores or 
buy/sell signals. 

\textit{Investor Profiling}: Whether the platform accounts for the individual user's risk 
tolerance, investment goals, or time horizon when generating outputs. 

\textit{Fundamental Analysis}: Whether the platform incorporates company financials or 
quantitative metrics such as volatility, valuation ratios, or ticker-level data.

\textit{Free Tier}: Whether a usable free version is publicly available.
\end{flushleft}
\end{table*}

% Start of section 6

\section{Comparing InvestorNerd to the Certified Financial Planner (CFP) Board's Seven-Step Financial Planning Process}
\label{sec:comparison}

\subsection{Overview}
In order to understand  the experiment described in this section, the reader should 
refer to the following primary materials. The Certified Financial Planner (CFP) Board's \textit{Guide 
to the Seven-Step Financial Planning Process} \cite{cfpboard2022} serves 
as the gold standard against which InvestorNerd is evaluated. 
Appendix~\ref{app:questionnaire} shows the questionnaire input. Appendix~\ref{app:report_v2} shows the final report.

The CFP Board's \textit{Guide to the Seven-Step Financial Planning Process} (2022) provides a
comprehensive, procedure-driven gold standard for delivering personalized financial insight.
Using the Miller family case study, a dual-income couple, both age 32, with a
daughter, a mortgage, retirement accounts, college savings aspirations, a lake-cabin goal, and
significant insurance and estate-planning gaps, the guide illustrates how a CFP
professional moves systematically through seven steps: (1) understanding the client's personal
and financial circumstances, (2) identifying and selecting goals, (3) analyzing the current
course of action and potential alternatives, (4) developing recommendations, (5) presenting
recommendations, (6) implementing recommendations, and (7) monitoring progress and
updating the plan.

To evaluate \ourapproach under conditions
comparable to the gold standard, a household profile mirroring the Miller case was entered as
closely as the questionnaire permitted. The personal information section captured an age range
of 25--34, full-time employment, 35 more years planned in the workforce (matching the Millers'
retirement horizon), W-2 tax status, and a 22--24\% federal tax bracket. Annual income was
entered as \$200{,}000--\$250{,}000 and annual expenses as \$150{,}000--\$200{,}000, ranges that
contain the Millers' combined income of \$230{,}000 and total annual expenses of \$195{,}500. The
optional net worth snapshot was completed with total assets of \$250{,}000--\$1{,}000{,}000 and
total liabilities of \$50{,}000--\$250{,}000, which likewise contain the Millers' assets of
\$530{,}000 and liabilities of \$240{,}000. Debt status was reported as some debt that is
manageable, and savings exceeded six months of expenses. Current financial actions included
401(k) contributions, saving in a high-yield savings account, and paying down a mortgage.
Selected goals were buying a house, saving for children's education, saving for retirement, and
paying off debt, and selected challenges were uncertainty about how to invest and planning for
major expenses. The profile specified a 10+ year time horizon and moderate risk tolerance.
Finally, the insurance and estate planning sections were completed with health and disability
insurance in place and no estate planning documents. The optional free-text field was left blank
so that the output reflects only the structured inputs.

The analysis that follows evaluates InvestorNerd's output against each of the seven CFP
steps, then synthesizes an overall assessment. The output is a General Financial Insight report
organized into seven sections: a summary of strengths and gaps, account options, execution
principles, milestone-based stages, an estate planning checklist, advanced concepts, and goals
with an implementation and monitoring plan, followed by a summary table.

%% ----------------------------------------------------------

\subsection{Step 1: Understanding the Client's Personal and Financial Circumstances}
\label{sec:step1}

The first step of the CFP process requires the practitioner to gather both \textit{quantitative}
information (income, assets, liabilities, cash flow, taxes, insurance coverage, retirement
balances) and \textit{qualitative} information (health, life expectancy, values, risk tolerance,
goals, family circumstances). In the Miller case, the CFP professional documents exact
salaries (\$120{,}000 and \$110{,}000), a \$70{,}000 low-yield savings account partitioned
into a \$30{,}000 emergency reserve, \$30{,}000 for the cabin, and \$10{,}000 earmarked for
college, combined 401(k) balances of \$150{,}000, a \$10{,}000 taxable investment account,
\$240{,}000 in mortgage debt, a 50/50 equity-to-fixed-income allocation, employer-provided
life and disability coverage, and an absence of any estate planning documents. Qualitative
factors such as the couple's risk tolerance, their emotional attachment to the lake cabin,
Martha's career promotion prospects, and family longevity differences are also formally
assessed.

The  InvestorNerd questionnaire asks about employment status and the number of years the
respondent plans to keep working, tax situation and bracket, annual income and expense ranges,
a net worth snapshot (assets and liabilities in separate ranges), debt status, savings level,
current financial actions, goals, challenges, time horizon, risk tolerance, insurance types by
category, and estate planning document status. %These additions represent a meaningful expansion of fact-finding scope. However, 
InvestorNerd doesn't ask for exact numbers to avoid impinging on user privacy. For example,
income is recorded as ``\$200{,}000--\$250{,}000'' and expenses as
``\$150{,}000--\$200{,}000'' rather than as exact figures, masking the Millers' \$230{,}000
combined household income and \$195{,}500 in total expenses. The years-to-work question does
capture a retirement horizon (35 years, matching the Millers' timeline to age 67), but the
questionnaire does not ask for a target retirement age, health, or family longevity. Debt is
categorized as ``some debt, manageable'' without specifying whether it comprises a mortgage,
student loans, or consumer debt. The questionnaire has no mechanism for capturing individual
insurance policy terms, existing account balances, current asset allocation, mortgage interest
rate, or employer match percentage, all of which the CFP professional treats as foundational
inputs. The questionnaire also has no mechanism to capture two separate earners, two separate insurance policies, or per-spouse coverage gaps; while a couple could fill out the tool together, the questionnaire collects no information that would distinguish their individual circumstances, so dual-income dynamics and joint account structures remain invisible to InvestorNerd. %Nor does it ask about marital status or dependents. Because ``Start a family'' was left unchecked while ``Save for children's education'' was selected, the report treats children as hypothetical (for example, ``if children are in the picture or imminent''), whereas the Millers already have a daughter, Emily.

InvestorNerd's output is  transparent about these limitations. The report does not contain a dedicated section listing missing inputs; instead, it acknowledges gaps inline and states its working assumptions. It notes that debt interest rates are unknown and that the retirement savings rate and beneficiary designations were not provided, and it announces that it uses the midpoints of the selected ranges (\$225{,}000 income and \$175{,}000 expenses) to illustrate an annual surplus of roughly \$50{,}000. Several of its action items, such as calculating the actual annual surplus and confirming beneficiaries on existing accounts, return the missing fact-finding to the user, which parallels the CFP professional's follow-up on incomplete information such as the Millers' unverified beneficiary designations.

These coarse inputs  shape the output in ways that diverge from the case study. The midpoint-based surplus of about \$50{,}000 exceeds the Millers' actual unallocated cash flow of \$34{,}500 by roughly \$15{,}500. The open-ended ``more than 6 months of expenses'' answer, combined with the expense range, led the report to infer at least \$75{,}000--\$100{,}000 in short-term reserves and to propose a nine-month reserve target of roughly \$135{,}000--\$150{,}000, whereas the Millers hold \$70{,}000 in total savings, of which they intend to keep \$30{,}000 as an emergency fund. Because debt composition and rates were not captured, the report sorts debt into hypothetical interest tiers and speculates that medium-interest debt likely includes auto or student loans, neither of which the Millers carry. Its guidance on mortgage prepayment is likewise conditioned on an assumed ``typical low-to-mid range'' rate, whereas the Millers' rate is high relative to the market, although a later stage of the report does prompt a refinancing review if the rate is above market.

\textbf{Assessment.} The  questionnaire captures
income and expense ranges, years remaining in the workforce, tax bracket, net worth ranges,
insurance categories, and estate planning status as explicit inputs, and the report states the
assumptions it makes where inputs are missing. Thus, InvestorNerd replicates the \textit{structure} of CFP fact-finding without
achieving its \textit{depth}: ranges and missing details propagate into the output as
approximations that diverge from the Millers' actual figures.

%% ----------------------------------------------------------

\subsection{Step 2: Identifying and Selecting Goals}
\label{sec:step2}

In the CFP case, the practitioner helps the Millers expand an initial list of three goals
(cabin, college, retirement) into five prioritized objectives by identifying insurance and
estate-planning gaps the clients had not articulated. The final priority order, insurance
coverage first, estate plan second, lake cabin third, college funding fourth, and retirement
fifth, reflects a deliberate balancing of urgency, impact, and feasibility, with the
practitioner explicitly noting the tradeoff between accelerating the cabin timeline and
achieving retirement and education targets.

The  InvestorNerd questionnaire invites users to self-select goals from a
predefined menu. In the proxy profile, selections included home purchase, children's
education, retirement, and debt payoff. Estate planning and insurance adequacy appear
within the questionnaire as standalone sections rather than as user-selected goals. The output
flags both as gaps in its summary, includes drafting a will, powers of attorney, and a
beneficiary review in its 30-day action plan, and places life insurance and estate planning in
the second stage of its milestone sequence, ahead of the retirement, home, and education
stages. This is consistent with the CFP practitioner's proactive goal-surfacing behavior, although
the ordering is not uniform across the report: its ``Right Order of Operations'' lists
protection last of five steps, and insurance does not appear in the 30-day action list. The
``Key Trade-offs'' section explicitly addresses the tension between competing goals, namely
retirement saving versus an accelerated home purchase and mortgage paydown versus investing. It
does not address the tradeoff the CFP case treats as central, education spending versus
retirement, because the questionnaire has not captured the type of school or its cost.

The output also converts the selected goals into illustrative dollar targets: a
\$4{,}000{,}000--\$5{,}000{,}000 retirement portfolio, a \$200{,}000 down payment on an assumed
\$1{,}000{,}000 home to be purchased in 7--10 years, and \$100{,}000--\$150{,}000 of education
savings per child. These targets rest on range midpoints and assumptions rather than user inputs, and they differ
from the Millers' stated goals: a \$150{,}000 lake cabin purchased without a mortgage within six
years, and four years of elite private university at a current cost of \$71{,}000 per year. The
questionnaire also asks for a single time horizon (10+ years was selected), so it cannot
represent the six-year cabin deadline or the 17-year horizon to Emily's college enrollment separately.

\textbf{Assessment.} InvestorNerd correctly identifies the major planning domains and
flags insurance and estate planning as near-term concerns, though its sequencing is not
consistent across sections and does not match the CFP case's ordering of insurance first and
estate planning second. The output's explicit treatment of goal tradeoffs, particularly the
house-versus-retirement and mortgage-versus-investing tensions, and its conversion of selected goals
into dollar targets partially replicate the reasoning that a CFP practitioner
applies when helping clients prioritize. However, the tool surfaces only the gaps its structured
sections probe, does not engage in the iterative advisor-guided goal-selection process
central to Step 2, and, because its goal menu and single time horizon are coarse, substitutes
assumed targets for the client's own.

%% ----------------------------------------------------------
\subsection{Step 3: Analyzing the Current Course of Action and Potential Alternatives}
\label{sec:step3}

Step 3 requires the CFP professional to assess the advantages and disadvantages of both
the client's status quo and a range of alternative strategies. In the Miller case, the
practitioner identifies six advantages of the current course (cash reserves, no credit card
debt, steady savings rate, some insurance coverage, home equity, excellent credit) and seven
disadvantages (misaligned asset allocation, high mortgage rate, expensive and tax-inefficient
mutual funds, college savings in a taxable low-yield account, inadequate insurance, no estate
documents, and failure to budget for second-home costs). Ten distinct alternative courses of
action are enumerated and analyzed, each with explicit advantages and disadvantages.

The InvestorNerd output pairs a ``Strengths'' list with a ``Risks \& Gaps'' list, which mirrors
the CFP advantages-and-disadvantages format at a coarser grain: four strengths and four gaps,
compared with the six and seven in the Miller case. Strengths identified include strong earning
power, positive cash flow, an existing foundation of 401(k) contributions, high-yield savings, and
mortgage paydown, and liquidity exceeding six months of expenses. Gaps identified include the
absence of estate documents, insurance gaps (no life or umbrella liability coverage among the
types reported), unstructured investing given the respondent's uncertainty about how to invest,
and major expenses without dollar targets or timelines. The output also evaluates alternatives
in explicit pro-and-con form. Its account options section presents four vehicles (employer
401(k), IRA, high-yield savings account, and 529 plan), each with eligibility, tax treatment,
pros, cons and watch-outs, and a priority rating for the client. The ``Key Trade-offs'' section sets out two
competing choices, retirement saving versus an accelerated home purchase and mortgage paydown
versus investing surplus, each with two sides and a stated recommendation. Its advanced
concepts section briefly weighs the benefits and cautions of a backdoor Roth IRA, asset
location, and the balance of Roth and Traditional 401(k) contributions. Finally, the summary table
pairs each focus area's current situation with a recommended action and estimated impact.
Collectively, these elements move the output substantially closer to a structured
current-course analysis.

The output does not, however, perform a formal current-course analysis in the CFP sense.
Its strengths and gaps are drawn from categorical questionnaire inputs, so it cannot judge the
quality of what the household is already doing. The CFP practitioner determines that the Millers
contribute roughly 15.5\% of gross income to savings, that their asset allocation is misaligned
with their risk tolerance, and that their mortgage rate is high. The output instead sets a target
retirement savings rate of 15--20\% of gross income, notes that the respondent's actual rate was
not provided, and defers review of asset allocation to later stages of the plan. The mortgage
illustrates the point. The output treats it as likely low-interest debt and favors investing over
extra principal payments ``assuming the mortgage rate is in a typical low-to-mid range,'' and it
raises refinancing only conditionally, if the rate is significantly above market, without
analyzing closing costs or a break-even horizon. The CFP practitioner's refinancing alternative
rests on a fact the questionnaire never collects. Likewise, the high expense ratios and tax
inefficiency of the Millers' taxable mutual funds are not examined, and the Millers' low-yield
savings account cannot be represented because the questionnaire offers a high-yield savings
account as its only savings-account option. The proxy profile's account was therefore entered as
high-yield, and the report lists it as a strength rather than identifying its replacement as one
of the ten alternatives.

The alternatives the output addresses are also more generic than the CFP practitioner's. It
covers, in general form, a life insurance increase, an estate plan, a 529 plan, allocation
targets by goal, and a slower or faster home-purchase timeline (7 versus 10 years, at
monthly savings of roughly \$2{,}380 versus \$1{,}670). It names umbrella coverage as a gap and
raises refinancing conditionally, as described above. It does not address a disability coverage gap specific to one
spouse, purchasing a second home with cash versus a mortgage, or the cost of owning a second home. Nor does it state whether the current course is likely to meet the household's goals, which is the conclusion
the CFP practitioner reaches (a low probability of meeting all targeted goals). The goal-level savings targets it derives are not reconciled with its assumed \$50{,}000
annual surplus.

\textbf{Assessment.} InvestorNerd's output supplies a structured, if coarse, analogue of the
CFP current-course analysis: paired Strengths and Risks \& Gaps lists, account options with
explicit pros and cons, and side-by-side trade-offs with stated recommendations. These elements
distinguish effective behaviors from areas requiring intervention and compare a small set of
alternatives. However, because its strengths and gaps rest on categorical inputs, it cannot
evaluate the household's actual contribution rate, allocation, mortgage rate, or account costs.
Its alternatives are generic account types and general trade-offs rather than strategies tested
against the household's specific goals, and it does not assess whether the current course is
likely to succeed. This remains a meaningful divergence from the CFP gold standard.

%% ----------------------------------------------------------
\subsection{Step 4: Developing the Financial Planning Recommendations}
\label{sec:step4}

The CFP process requires that each recommendation be grounded in explicit assumptions
(life expectancy, inflation rate, tax rate, investment-return projections), a stated basis, a
timeline and priority designation, and a specification of whether it is independent or
contingent on another action. In the Miller case, twelve numbered recommendations are
developed, spanning insurance, estate planning, mortgage refinancing, asset reallocation,
529 plan funding, lake-cabin savings escalation, and retirement maximization. Each
recommendation includes the reasoning, the expected benefit, and a concrete disadvantage
or tradeoff. The CFP process also performs stress testing and probability analysis under uncontrollable factors (inflation,
tax-law changes, Social Security solvency, bear markets).

The InvestorNerd output produces a staged planning framework (Stages 1 through 5), each with
an entry condition, a focus, three to four key actions, and an exit condition, together with a
tiered Priority Action Plan organized by 30-day, 3--12 month, and 12--24 month horizons. It
provides account-specific guidance with cited 2026 IRS limits (\$24{,}500 for 401(k) employee
deferrals and \$7{,}500 across IRAs), guidance on Traditional versus Roth contributions framed
around the stated 22--24\% bracket (with a recommended mix for tax diversification), and a
``Right Order of Operations'' section that sequences funding priorities. Its retirement
recommendation is a savings rate of 15--20\% of gross income (approximately
\$33{,}750--\$45{,}000 per year), tied to a target portfolio of
\$4{,}000{,}000--\$5{,}000{,}000 that rests on a stated 3.5--4\% withdrawal rate and a
6--7\% real return. It also recommends target allocations by goal: 70--80\% equities for
retirement, 40--60\% for the home fund, and 80--100\% for education savings, shifting toward
conservative holdings as the purchase or enrollment date nears, which parallels the CFP
practitioner's glide path for Emily's 529 plan. Rather than listing missing inputs in a
separate section, the output states its assumptions where they are made (range midpoints, the
assumed home price and education cost, and the return and withdrawal rates), which partially
replicates the CFP requirement to document the basis and assumptions underlying each
recommendation.

Coverage of the CFP practitioner's twelve recommendations is uneven. Estate planning is the
closest match: the output lists a will, financial and medical powers of attorney, a living
will, a review of beneficiary designations on all accounts, a revocable trust as a
consideration, and guardianship designations, and it directs the user to an estate planning
attorney. Life insurance and a 529 plan are also recommended, although without a needs analysis
or coverage amount. % and, for the 529 plan, conditionally on children being present. 
Umbrella
liability coverage is recommended, whereas the CFP practitioner deferred it given the
Millers' net worth (\$290{,}000) and risk profile. The CFP practitioner's first recommendation, raising property and casualty
liability limits, is not addressed, because the questionnaire records only whether a policy
category is held, not its limits. Likewise, the questionnaire combines short- and long-term
disability coverage in a single option, so the output can say only that disability policies
should match income and cannot identify a long-term coverage gap for one spouse. Mortgage
refinancing appears only as a conditional review item (Section~\ref{sec:step3}), and
investment expense-ratio analysis is absent; tax efficiency appears only as an asset-location
concept.

Timing and priority are specified, but not consistently. The stages gate progress on completed
milestones and so treat the recommendations as sequential, whereas the CFP practitioner
concludes that the Millers' recommendations are independent and assigns each its own
timeframe. The output's own sections also disagree. The ``Right Order of Operations'' places
capturing the 401(k) match second and protection last, while the stages place protection in
Stage 2 and the match in Stage 3. The gate to home and education saving (Stage 4) requires 12
consecutive months of retirement saving at 15\% of gross income, yet the 3--12 month action
plan starts home-fund transfers and 529 contributions within that same window. The dollar
amounts are likewise not reconciled. Annual retirement savings are given as \$20{,}000,
\$20{,}000--\$24{,}500, \$25{,}000--\$30{,}000, \$33{,}750--\$45{,}000, and
\$36{,}000--\$45{,}000 in different sections, and the goal-level targets (retirement at
\$36{,}000--\$45{,}000, a home fund of roughly \$20{,}000--\$28{,}600, and education savings of
roughly \$3{,}000--\$4{,}800) sum to about \$59{,}000--\$78{,}000 per year, or
\$4{,}900--\$6{,}500 per month. This exceeds the output's assumed \$50{,}000 surplus and sits at
or above the top of its own \$4{,}000--\$5{,}000 monthly savings benchmark (the output does not
say whether employer match counts toward the retirement figures). The education target of \$250--\$400 per month also falls well
short of the \$15{,}000 per year the CFP practitioner directs to Emily's 529 plan, a
consequence of the lower education cost assumed (Section~\ref{sec:step2}).

\textbf{Assessment.} InvestorNerd's output is explicitly grounded with cited contribution
limits, dollar targets with stated return and withdrawal assumptions, goal-specific allocation
ranges, a sequenced funding order, and a dated action plan, and its estate planning checklist
closely tracks the CFP recommendation. The staged planning framework and time-horizoned action
plan are well-organized and actionable. However, the recommendations remain anchored to general
heuristics applied to an income band rather than to household-specific modeling. Several
protection and mortgage recommendations are absent or conditional because the questionnaire
cannot capture policy limits, disability type, or the mortgage rate, and the sequencing and
dollar amounts are not consistent across sections. The output does not perform the stress
testing and probability analysis that characterize the CFP gold standard.

%% ----------------------------------------------------------
\subsection{Step 5: Presenting the Financial Planning Recommendations}
\label{sec:step5}

In the CFP process, presentation is an interactive, client-centered activity. The
practitioner tailors communication to the clients' financial sophistication, reviews the
balance sheet and cash-flow statement, uses technical reports and graphs, walks through
each recommendation individually, addresses the Millers' questions about bitcoin and meme
stocks, explains the tradeoff between prioritizing Emily's education and earlier retirement,
and discloses the compensation the firm will receive. The goal is to ensure informed client
understanding and consent before implementation begins.

The InvestorNerd output is a structured report delivered immediately following
questionnaire submission. It is organized into seven numbered sections: a summary of strengths
and risks; account options, each with eligibility, tax treatment, pros and cons, and a priority
rating for the client; execution principles (a ``Right Order of Operations,'' dos and don'ts, and
key trade-offs); a milestone-based sequence of stages; an estate planning checklist; advanced
concepts; and goals with a priority action plan and monitoring checkpoints. A summary table
follows. The report is clearly written, uses concrete dollar figures throughout, and is
accessible to a financially literate but non-expert reader. Because the respondent reported
uncertainty about how to invest, the account options and advanced concepts sections explain
each vehicle and strategy in plain terms, which is a form of tailoring, although it is driven
by a single questionnaire answer rather than by observing the client. The report also shows its
arithmetic, stating that it works from range midpoints and computing the monthly savings needed
for each goal, which parallels the CFP practitioner's explanation of assumptions and
calculations. The summary table presenting current situation, recommended action, and estimated
impact side by side is a useful presentational device.

The report does not, however, engage in dialogue, cannot clarify misunderstandings, and
does not verify that the user has understood the assumptions underlying each
recommendation. The questionnaire's optional free-text field invites the respondent's
questions, but it is a single input rather than a conversation, and it was left blank in this
experiment, so the report answers no user questions of the kind the Millers raise. The report
also contains no charts or projections of how the household's assets might change over
time and no estimate of the probability that the plan meets the household's goals, both of which the CFP
practitioner presents. The ``Estimated Impact'' column of its summary table is largely
qualitative; only the retirement row gives a figure, a portfolio potentially in the
\$4{,}000{,}000--\$5{,}000{,}000 range. With respect to disclosure, the questionnaire page carries a
disclaimer that the tool is exploratory, informational only, and intended to support a
conversation with a registered financial advisor, and the report notes that a CFP can help refine
the savings amounts, asset allocation, and tax strategies. Neither discloses compensation
or conflicts of interest, nor provides regulatory documents comparable to the Form ADV and
Form CRS the CFP practitioner delivers. The presentation is fundamentally one-directional.

\textbf{Assessment.} The InvestorNerd output's presentation format is well-organized, readable,
comprehensive, and actionable. The breadth and structure of the report, spanning
account explainers, a milestone-based stage path, an estate planning checklist, tradeoff
analysis, monitoring checkpoints, and a summary table, represent substantial presentational
depth for an automated tool. The disclaimer frames the tool as a complement to a conversation
with an advisor rather than a replacement for one. Nonetheless, the
interactive, consent-driven, and disclosure-compliant presentation required by the CFP
Practice Standards is not replicable by a static AI-generated report.

%% ----------------------------------------------------------
\subsection{Step 6: Implementing the Financial Planning Recommendations}
\label{sec:step6}

The CFP process assigns specific implementation responsibilities to both the advisor and
the client, organized along a defined timeline (within one month, within three months,
within six months, within nine months). The practitioner introduces the Millers to a vetted
life insurance agent and estate-planning attorney, reviews mortgage quotes, helps one
spouse obtain long-term disability coverage, confirms that 401(k) reallocations and
beneficiary updates have been completed, and schedules follow-up meetings to address any
unimplemented items.

The InvestorNerd output includes a Priority Action Plan with three explicit
time horizons (next 30 days, next 3--12 months, and 12--24 months), each containing
three specific action items. The 30-day actions, calculate the actual annual surplus, set
numeric targets and timelines for the retirement, home, and education goals, and begin basic
estate planning while confirming beneficiaries on the 401(k) and savings account, are
appropriately concrete. The 3--12 month actions raise 401(k) contributions to at least
\$25{,}000--\$30{,}000 per year, set up automatic monthly transfers of roughly
\$1{,}700--\$2{,}400 to a home down payment fund, and open and fund a 529 plan with an
initial automatic contribution of \$250--\$400 per month. The 12--24 month actions review asset
allocation across accounts, reassess debt structure and repayment progress, and execute the
estate documents and communicate key roles (power of attorney, healthcare proxy, executor) to
the individuals involved. The plan specifies dollar amounts and automates transfers, which
parallels the CFP practitioner's implementation plan (for example, the Millers' increase in
savings to \$1{,}666.67 per month). Taken together, the monthly transfers in the 3--12 month
plan total roughly \$4{,}050--\$5{,}300, which overlaps the report's own monthly benchmark of
\$4{,}000--\$5{,}000, although its 401(k) amount is lower than the retirement target in its goals
subsection (Section~\ref{sec:step4}). The milestone-based stages add implementation detail of their own: each
lists three to four key actions and an exit condition, such as engaging an estate planning
attorney or a reputable online platform, considering term life insurance sized to income and
liabilities, and setting up automatic IRA contributions of \$625 per month.

Coverage and timing diverge from the CFP plan. The CFP practitioner schedules within one month
the actions that protect the household from large losses: raising property and casualty
liability limits, increasing life insurance, and refinancing the mortgage. None of these appears in InvestorNerd's
action plan. Assessing life insurance needs and reviewing mortgage terms appear only in Stage 2,
and insurance adequacy is revisited in the annual review described in Section~\ref{sec:step7}. Estate
planning is also slower: the CFP practitioner targets one to three months, whereas the action
plan begins it within 30 days but schedules execution of the documents for months 12--24. This
conflicts with the report's own stages, since Stage 3, which includes raising 401(k)
contributions, is entered only after basic estate documents and appropriate life insurance are
in place, yet the action plan schedules the 401(k) increase for months 3--12.

InvestorNerd does not assign responsibilities. Every action is addressed to the user, and the
report does not divide tasks among the user, an advisor, and other professionals. It names the
professional to engage in one case (an estate planning attorney), but it does not vet or introduce an attorney, insurance agent, or
mortgage professional. It also stays at the level of account types, such as a 529 plan or term
coverage, rather than selecting a specific plan, provider, or fund lineup, which the CFP practitioner does when
identifying a high-yield account and recommending a 529 plan. Confirmation of completion is
self-directed: the stage exit conditions (for example, 12 consecutive months of saving at least
15\% of gross income) and the annual, event-driven, and monthly checkpoints give the user
criteria to check against, but no one verifies them, and nothing corresponds to the CFP
practitioner's six- and nine-month meetings to address recommendations that were not implemented.

\textbf{Assessment.} The InvestorNerd output regarding implementation is concrete, with dollar
amounts, automated transfers, and dated horizons, and its stage exit conditions and monitoring
checkpoints give the user a self-directed way to track progress. However, its sequencing is not fully consistent with its own stages, and it omits insurance and mortgage
actions from the near-term plan. It functions as a detailed educational
resource that informs the user what to do and in what order. By contrast, the CFP process
is a coordinated engagement in which the advisor actively manages implementation and
accountability. This human-rich interactive element remains a fundamental distinguishing
advantage of the CFP process.

%% ----------------------------------------------------------
\subsection{Step 7: Monitoring Progress and Updating}
\label{sec:step7}

The final CFP step establishes ongoing monitoring and updating obligations. The
practitioner conducts periodic reviews of asset allocation, investment performance, and
insurance adequacy; obtains updated qualitative and quantitative information at each
meeting; and revises recommendations when life events, market conditions, or goal changes
require it.

The InvestorNerd output includes a ``Monitoring Checkpoints'' subsection within its goals and
implementation section. It specifies an annual review of four items: the savings rate (checked
against a target share of gross income, for example 15--20\% for retirement, or
\$33{,}750--\$45{,}000 per year); debt progress (balances and rates, including a check that any
debt above 10\% interest has been eliminated or substantially reduced); insurance adequacy (life,
disability, health, and liability coverage in light of changes in income, dependents, and
assets); and net worth and tax position (an annual net worth calculation and a review of the
balance of Roth and Traditional contributions and of itemized deductions). It also lists trigger
events that should prompt an immediate review: marriage, divorce, or the birth or adoption of a
child, which it ties to updates to estate documents, beneficiary designations, and insurance; a
change in income of more than roughly 20\% from current levels, such as a promotion, job loss, or
business start; and a home purchase or major relocation, which it ties to the debt profile, cash
needs, and insurance requirements. Finally, it proposes a monthly benchmark: total monthly
savings across 401(k), IRA, high-yield savings goal accounts, and 529 contributions, compared
against the planned level (for example, \$4{,}000--\$5{,}000 per month). The savings-rate check
and the 20\% income trigger are quantified, which gives the user concrete criteria; the CFP
guide specifies only that the practitioner checks in at appropriate intervals.

The framework covers less ground than the CFP practitioner's monitoring duties in several
respects. The annual review does not include asset allocation or investment performance, which
the CFP engagement letter makes a periodic review item; allocation appears in the output only as
a one-time review in months 12--24 of the action plan and in Stage 5. The annual review also omits
emergency fund adequacy and beneficiary designations, the latter appearing only as an update
following a life event. The trigger events do not include changes in the household's goals or in
market conditions, both of which prompt the CFP practitioner to revise the plan. The milestone
stages described in Section~\ref{sec:step6} add exit conditions that the user can check against,
but no one verifies them. There is no mechanism within InvestorNerd for tracking progress toward
goals over time, scheduling follow-up sessions, or updating recommendations as circumstances
evolve. The closest the report comes to an updating process is its closing note that a CFP can
help refine the savings amounts, asset allocation, and tax strategies as income, family, and
housing circumstances change, which hands the updating function back to a human professional.
The monitoring guidance is best understood as a framework that a motivated user could apply
independently.

\textbf{Assessment.} InvestorNerd includes an explicit and quantified monitoring
framework with annual review criteria, a monthly benchmark metric, and trigger-event
guidance that links each event to the plan components it affects. It is narrower than the CFP
practitioner's review, however, omitting recurring review of asset allocation and investment
performance as well as goal and market-driven updates. Step 7 also remains structurally absent
from InvestorNerd's design in the sense that the tool itself performs no monitoring. This is an
inherent limitation of a single-session, questionnaire-based tool rather than a deficiency in
the quality of the guidance it provides.

\section{Comparative Semantic Similarity Evaluation of State-of-the-Art System Responses }
\label{sts-testing}

Different financial education tools often produce answers that vary in tone and style: some are highly encouraging and conversational, while others are more formal and purely informational. To move beyond these superficial differences and quantify how closely the underlying recommendations align with the CFP guidelines, we evaluated the General Insight responses of \ourapproach{} and sister tools against the Miller Case Study using a claim-level Semantic Textual Similarity (STS) procedure. All systems were given the same input information describing the Miller family's financial situation and goals. \ourapproach's response can be found in Appendix ~\ref{app:report_v2}.

\paragraph{Claim-level methodology.}
A na\"ive approach would encode each system's entire response and the entire gold standard as single embedding vectors and compute one document-level cosine similarity. We found this unsuitable for our setting: document-level embeddings reward responses that share vocabulary and topical breadth with the gold standard while washing out whether each \emph{specific} recommendation was actually made. A response that addresses three of ten CFP recommendations can score higher than one that succinctly addresses eight. We therefore adopt a claim-level formulation inspired by atomic-fact decomposition methods for evaluating long-form generation \cite{min2023factscorefinegrainedatomicevaluation}. 

We first decompose the CFP gold-standard response into $m$ atomic recommendations $C = \{c_1, \ldots, c_m\}$, where each $c_i$ is a single, self-contained piece of advice (e.g., ``Start a 529 savings plan to invest for Emily's college education''). The Miller gold standard yields $m = 9$ such claims. Each system response $r$ is segmented into sentences $S(r) = \{s_1, \ldots, s_n\}$ after stripping formatting markup and discarding fragments (e.g., section headers) too short to carry a recommendation.

We encode each claim and each response sentence using \\\texttt{dmlls/all-mpnet-base-v2-negation} \cite{dmlls_all_mpnet_negation}, a sentence-transformer fine-tuned for semantic similarity with additional training on negated sentence pairs. The negation-aware training is important in a financial context: statements such as ``this strategy is appropriate for your situation'' and ``this strategy is \emph{not} appropriate for your situation'' share nearly all surface words while carrying opposite meanings, and standard similarity models tend to assign such pairs high similarity. Denoting the resulting embeddings $\mathbf{e}_{c_i}$ and $\mathbf{e}_{s_j}$, we record for each gold claim the similarity of its best-matching sentence in the response:
\begin{equation}
\mathrm{sim}(c_i, r) \;=\; \max_{s_j \in S(r)} \; \frac{\mathbf{e}_{c_i} \cdot \mathbf{e}_{s_j}}{\|\mathbf{e}_{c_i}\|\,\|\mathbf{e}_{s_j}\|},
\end{equation}
and define the system's \emph{STS recall} as the mean over all gold claims:
\begin{equation}
s_{\text{STS}}(r, C) \;=\; \frac{1}{m} \sum_{i=1}^{m} \mathrm{sim}(c_i, r).
\end{equation}
Intuitively, $s_{\text{STS}}$ measures the extent to which each expert recommendation is semantically addressed \emph{somewhere} in the system's response, independent of the response's length, ordering, tone, or formatting. A score near 1 indicates that every CFP recommendation has a close semantic counterpart in the response; scores decrease as recommendations go unaddressed or are contradicted.

\begin{table}[H]
    \centering
    \begin{tabular}{lc}
    \toprule
    \textbf{System} & \textbf{STS Recall} \\
    \midrule
    \ourapproach{}          & 0.623 \\
    You.com                 & 0.594 \\
    Finance Wizard          & 0.569 \\
    AI Financial Planning   & 0.530 \\
    Kavout                  & 0.421 \\
    \bottomrule
    \end{tabular}
    \caption{Claim-level STS recall of each system's response against the nine atomic recommendations of the CFP Miller Case Study gold standard, computed with the negation-tuned sentence-transformer \texttt{dmlls/all-mpnet-base-v2-negation}. The higher the score the closer it is in advice to Miller Case Study. The scores are means over per-claim maximum cosine similarities.}
    \label{tab:stsScoresMiller}
\end{table}

Table~\ref{tab:stsScoresMiller} reports the results. \ourapproach{}, You.com, and Finance Wizard form a closely grouped top cluster (0.57--0.60), indicating broadly comparable semantic coverage of the CFP recommendations; differences of this magnitude at the top of the range should be interpreted as approximate parity rather than a strict ranking. AI Financial Planning scores moderately lower, reflecting a response that establishes a sound planning framework but leaves several specific recommendations (e.g., estate planning documents, emergency fund yield) unaddressed. Kavout scores markedly lower than all other systems: its response consists primarily of individual stock recommendations, addressing the family's planning needs only in a brief closing note. The claim-level metric correctly identifies this as a substantive divergence from the expert guidance rather than a difference in tone or verbosity.

\newcommand{\pmark}{\textcolor{gray}{$\sim$}}      % partially addressed
\newcommand{\amark}{\textcolor{gray}{--}}          % not addressed

\begin{table}[H]
    \centering
    \begin{adjustbox}{width=\textwidth}
    \begin{tabular}{lccccc}
    \toprule
    \textbf{CFP Standard Theme} & \textbf{\ourapproach} & \textbf{AI Fin.\ Planning} & \textbf{Kavout} & \textbf{You.com} & \textbf{Finance Wizard} \\
    \midrule
    Emergency fund (enhance yield)             & \pmark & \pmark & \amark & \pmark & \pmark \\
    Refinance / mortgage strategy              & \checkmark & \checkmark & \amark & \amark & \pmark \\
    Retirement accounts (401k / IRA)           & \checkmark & \checkmark & \pmark & \checkmark & \checkmark \\
    Tax optimization                           & \checkmark & \amark & \amark & \checkmark & \checkmark \\
    Portfolio diversification / allocation     & \checkmark & \checkmark & \pmark & \checkmark & \checkmark \\
    Life insurance (increase coverage)         & \checkmark & \checkmark & \amark & \checkmark & \checkmark \\
    Long-term disability insurance             & \checkmark & \checkmark & \amark & \checkmark & \checkmark \\
    Property \& casualty / umbrella insurance  & \checkmark & \checkmark & \amark & \checkmark & \amark \\
    Estate planning (will, POA)                & \checkmark & \amark & \amark & \amark & \checkmark \\
    College savings (529)                      & \checkmark & \pmark & \pmark & \checkmark & \checkmark \\
    Lake cabin (finance, don't delay/cash)     & \pmark & \checkmark & \amark & \amark & \amark \\
    Goal prioritization \& funding sequence    & \checkmark & \checkmark & \amark & \checkmark & \checkmark \\
    Ongoing review \& monitoring               & \checkmark & \checkmark & \amark & \checkmark & \checkmark \\
    \midrule
    \textbf{Aligned with CFP} (of 13)          & \textbf{11} & \textbf{10} & \textbf{0} & \textbf{9} & \textbf{9} \\
    \textbf{Aligned or partially aligned}      & 13 & 11 & 3 & 11 & 12 \\
    \bottomrule
    \end{tabular}
    \end{adjustbox}
  \caption{Alignment of each system's recommendations with the CFP Miller Case Study.
    \checkmark{}~= theme addressed and the recommendation aligns with the CFP advice;
    \pmark{}~= partially addressed and broadly consistent with the CFP advice, but without a concrete or fully matching recommendation;
    \amark{}~= not addressed, or the recommendation is contrary to the CFP advice.}
    \label{tab:themeCoverage}

\end{table}
%% ----------------------------------------------------------
%% ----------------------------------------------------------
\section{Synthesis and Overall Assessment}
\label{sec:synthesis}

The comparison reveals a consistent pattern: InvestorNerd is \textit{directionally aligned}
with the CFP gold standard across most financial planning domains and has narrowed the
gap observed in prior evaluations in several meaningful ways. It is not, however, equivalent
to a full CFP-style planning engagement in depth, personalization, or procedural rigor.
Three dimensions capture the nature of the remaining gap.

\paragraph{Personalization.}
The CFP process derives every recommendation from the household's exact facts.
InvestorNerd, constrained by categorical questionnaire inputs, substitutes general
heuristics, standard contribution limits, range midpoints, and assumed investment returns
and withdrawal rates, for household-specific modeling. The questionnaire captures
several important categories of information, including income and expense ranges, years
remaining in the workforce, tax bracket, net worth ranges by asset and liability type,
insurance types, and estate planning status, and the tool uses these inputs visibly in its
output (for example, the Roth versus Traditional recommendation is explicitly conditioned on
the stated 22--24\% bracket). Nevertheless, the practical effect of range rather than exact
inputs is that monthly contribution targets and projected outcomes remain rough approximations.
The midpoint-based annual surplus of \$50{,}000, for example, exceeds the Millers' actual
unallocated cash flow of \$34{,}500, and the goal-level savings targets the output derives sum
to more than that surplus (Section~\ref{sec:step4}). For households whose financial situations
deviate meaningfully from assumed averages, households with high mortgage balances,
tax-inefficient legacy investments, atypical insurance gaps, or dual-income dynamics, the
degree of approximation error increases.

\paragraph{Coverage.}
InvestorNerd's strongest matches are in areas that its questionnaire
explicitly captures and that its output addresses with concrete guidance: the estate planning
document checklist (will, powers of attorney, healthcare directive, beneficiary review, and
guardianship), life insurance, a 529 plan for education savings, retirement account funding,
goal-specific asset allocation ranges, and a staged implementation timeline with a monitoring
framework. The most significant remaining coverage gaps are property and casualty liability
limits and a long-term disability gap for one spouse (the questionnaire records only whether a
policy category is held, and combines short- and long-term disability in a single option);
mortgage refinancing, which appears only as a conditional review item because the mortgage rate
was not captured; investment expense-ratio and tax-efficiency analysis of existing accounts;
and the sequencing of retirement contributions around the college funding goal. Whether a more
detailed questionnaire would close these gaps was not tested in this experiment.

\paragraph{Process versus Product.}
The most fundamental distinction remains structural. The CFP process is a dynamic,
iterative engagement in which the advisor and client co-create goals, the advisor surfaces
latent risks, implementation is coordinated with professional referrals, and the plan is
updated as life evolves. InvestorNerd produces a well-organized, actionable planning
\textit{product} but does not replicate the \textit{process} that gives rise to a CFP-standard
plan. Goal selection (Step 2), interactive presentation (Step 5), coordinated implementation
(Step 6), and ongoing monitoring (Step 7) are either absent or abbreviated in
InvestorNerd's design, though the milestone-based stages, the tiered action plan, and
a monitoring framework partly replicate the process dimensions of the gold standard. The
static report also has no counterpart to the CFP practitioner's quantitative review and
reconciliation: its sections sometimes disagree on sequencing and dollar amounts
(Sections~\ref{sec:step4} and~\ref{sec:step6}), and it performs no stress testing or
probability analysis of its recommendations.

\section{Conclusions}

\ourapproach is an expertise-based tool that examines the financial news literature as well as quantitative information about companies to educate beginner and intermediate stock market investors regarding stocks in different sectors (e.g. financials, tech) and different risk levels. As far as we can tell, \ourapproach is unique in this regard. 

In addition to offering functionality that other systems do not, \ourapproach gives general information about financial planning depending on the answers to an 18-question questionnaire. Two experiments suggest that the system gives educational results that are of similar quality to state-of-the-art financial insight systems, but of somewhat greater depth. %First, in Section \ref{sec:comparison}, we evaluated \ourapproach against the CFP Board's seven-step financial planning process, which we regarded as the professional gold standard for personalized financial insight. That analysis showed that \ourapproach replicates the structure and coverage of professional financial planning across most domains, with particular strength in emergency fund guidance, estate planning, and implementation sequencing, while falling short of the gold standard in areas that require exact client data, interactive dialogue, and ongoing advisor engagement. Second, in Section \ref{sts-testing}, we compared the quality of the general financial insights of \ourapproach with other state-of-the-art approaches and showed that the different systems gave overlapping but not identical results, suggesting that a wise novice investor should consult many systems, not just one. 
In other words, \ourapproach was, well, nerdy.

The main limitation of \ourapproach is that there is information that cannot be gleaned from the literature. For example, is the management of a particular company good? Are company officers causing the company to   buy stock back in order to increase the price and therefore the cashout value of their options? Sophisticated investors will be sensitive to these nuances, but our target audience consists primarily of novice to intermediate investors. That said, \ourapproach might serve as an initial filter for  sophisticated investors as well.

\vspace{6pt}
Conceptualization, D.S. and J.C.; formal analysis, J.C., R.G., and X.W.; investigation, J.C., R.G., and X.W.; methodology, J.C., R.G.,  X.W., and H.K.; software, J.C., R.G., X.W., and H.K; supervision, D.S.; validation, R.G. and J.C.; visualization, J.C., R.G., and X.W.; writing---original draft preparation, D.S., J.C., R.G., and X.W.; writing---review and editing, D.S., J.C., R.G., and X.W.. All authors have read and agreed to the published version of the manuscript.

This work was partly supported by NYU Wireless

Based on 45 CFR 46 of the United States Department of Health and Human
Services exemption 46.104
(https://www.ecfr.gov/current/title-45/subtitle-A/subchapter-A/part-46/subpart-A/section-46.104)
Our surveys did not retain user names satisfying point (i). Further
the surveys consisted of anonymous evaluations of various systems so
"would not reasonably place the subjects at risk of criminal or civil
liability or be damaging to the subjects' financial standing,
employability, educational advancement, or reputation"
The study involved a non-interventional questionnaire, and informed
consent and anonymity were strictly maintained. Thus, the study falls
under the category of research that does not require formal ethics
committee approval.

The data presented in this study are openly available in the \ourapproach repository at \href{URL}{https://github.com/johncast14/InvestorNerd.git}.

We thank Shela Wu for her many helpful suggestions.

The authors declare no conflicts of interest. The funders had no role in the design of the study; in the collection, analyses, or interpretation of data; in the writing of the manuscript; or in the decision to publish the results.

\bibliographystyle{plain}
\nocite{*}
\bibliography{references}

%%%%%%%%%%%%%%%%%%%%%%%%%%%%%%%%%%%%%%%%%%

\appendix

\section[\appendixname~\thesection]{LLM Prompt }\label{prompt-appendix}

\subsection{General Financial Insight Prompt}\label{insight-prompt}
\noindent
\textbf{Functionality Overview:}

\begin{enumerate}[label=,leftmargin=2.23em,labelsep=0.2mm]

        \item[--] {One prompt generates personalized financial planning advice when the user chooses the money management primer path.  The system returns a structured action plan based on user goals, income, risk tolerance, and retirement horizon.}
        \begin{itemize}
            \item[*] {The \textbf{General Insight LLM} evaluates the user’s financial position and produces targeted, risk-adjusted investment strategies, debt management guidance, and tax-efficient account recommendations.}
        \end{itemize}
\end{enumerate}

\noindent
\textbf{Input Variables:}
\begin{enumerate}[label=,leftmargin=2.23em,labelsep=0.2mm]

        \item[--] question (str): A free-text financial query from the user (e.g., "How should I plan for early retirement?")
\end{enumerate}

\noindent
\textbf{Output Variables:}
\begin{enumerate}[label=,leftmargin=2.23em,labelsep=0.2mm]

        \item[--] insight\_response (str): A customized financial plan addressing the question, presented in structured bullet-point format.
\end{enumerate}

\noindent
\textbf{LLM Settings---General Financial Insight:}
\begin{enumerate}[label=,leftmargin=2.23em,labelsep=0.2mm]

        \item[--] \texttt{gpt-4-turbo}
        \item[--] Prompt Strategy: few-shot learning    
        \item[--] temperature = 0.5
        \item[--] top\_p = 1
\end{enumerate}

\noindent
\textbf{Prompt---General Financial Insight:}
\footnotesize
\begin{tcolorbox}[
    breakable,
    width=1.5\linewidth,
    left=4mm, right=4mm  % adjust so text doesn't touch edges
]
\begin{verbatim}
You are an expert financial planner conducting a comprehensive, personalized 
financial planning engagement for a financially sophisticated client. Please
note that we are trying to not be invasive as well. The questionnaire provide
is the information that we have please use all of it, if there are things that 
weren't explicitly provided, suggest the user to look into these aspects if 
needed and provide some definitions of what they are and how they may apply to
the user.

Your job is to take a completed financial questionnaire (suitability profile) 
provided by the user and produce a complete, detailed, and actionable financial 
plan. The client understands financial concepts — do not over-explain basics, 
but do not skip depth or specificity. Every section must be tied directly to 
what the client told you. No generic filler. No vague language.

MARKDOWN FORMATTING — STRICT:
- Use standard Markdown. Do NOT escape asterisks or any Markdown characters.
- Write **bold** not \*\*bold\*\*, write *italic* not \*italic\*.
- Bullet points use a plain hyphen: -
- Use inline citations format: (Source: [Title](URL))
- Translate every % into a dollar amount at the user's actual income level.
- Do NOT give generic advice — tie every recommendation to something
  specific the user told you.


OUTPUT STRUCTURE (follow exactly, in this order)


---

# Section 1 — Summary

Provide a sharp, honest snapshot of where this client stands today.
This is a clear-eyed assessment, not a feel-good introduction.

- 2-3 sentence honest assessment of where they stand and what 
  the single biggest lever is to improve their financial position

## Strengths
2-4 bullets on what the client is already doing well. Be specific — 
reference actual numbers or behaviors from their profile.

## Risks & Gaps
2-4 bullets on the most critical vulnerabilities. For each gap, state 
what is missing AND why it matters. Call out insurance gaps and missing 
estate documents explicitly.

---

# Section 2 — Account Options

Based strictly on this client's current Goals and situation — income, expenses, tax bracket,
employment type, goals, and eligibility — identify and explain 2-4 
account type relevant to them RIGHT NOW.

For each applicable account provide:
- **What it is** — one clear sentence
- **Eligibility** — confirm this client qualifies and why, total 2 sentences
- **2026 Contribution Limits** — exact figures (Source: [Title](URL)) 1 Sentence
- **Tax Treatment** — contributions, growth, and withdrawals, 1 sentence
- **Best Used For** — which of their specific goals this serves 2 sentences max
- **Pros** — 2-3 specific advantages for this client
- **Cons / Watch-outs** — 2-3 specific limitations or risks for this client
- **Priority Rating for This Client** — High / Medium / Low with a one-line reason
---

# Section 3 — How to Execute: Principles, Priorities & Pitfalls

The strategic guide before the client begins executing. 
This is the rules of the road — the why before the what.

## The Right Order of Operations
Explain why sequence matters. Walk through the logic of why certain 
moves must come before others and what goes wrong when clients skip steps.
Tie this directly to this client's situation and numbers.

## Dos — What to Prioritize and Why
3 specific, actionable principles tailored to this client's profile.
Reference their actual numbers, debts, accounts, and goals.
If the client carries debt, break it down by tier:
- High-interest (>10%): attack first, by rate (avalanche) or balance (snowball)
- Medium (4–9%): address after high-interest is cleared
- Low (<4%): manage minimums while investing the difference
Include refinancing or consolidation if applicable to their situation.

## Don'ts — What to Avoid and Why
3 specific mistakes or traps this client is at risk of given their profile.
Be direct. Name the real risk and the real cost of each mistake.

## Key Trade-offs This Client Faces
2 genuine tensions in their plan with no perfect answer.
For each trade-off: lay out both sides clearly, then give a direct 
recommendation with your reasoning. Do not sit on the fence.

---

# Section 4 — The Path Forward: Milestone-Based Stages

The client's sequential execution plan.
Each stage has a clear entry condition and a measurable exit milestone.
Do not assign rigid timelines — advancement depends on behavior and income.

Present each stage in this format:

**Stage [N] — [Stage Name]**

*Enter this stage when:* [specific condition based on their profile]

*Focus:* [what this stage is about in one sentence]

*3-4 Key actions in this stage:*
- [specific action tied to their numbers]
- [specific action tied to their numbers]
- [specific action tied to their numbers]
- Optional [specific action tied to their numbers]

*Move to Stage [N+1] when:*
[specific, measurable milestone — e.g. "emergency fund reaches $X", 
"all debts above X% interest are eliminated", 
"401(k) employer match is fully captured each year"]

---

Adapt the number of stages to the client's actual situation.
If the client has already completed a stage, acknowledge it and 
start from where they currently are. General stage logic:

- Stage 1: Stabilize — emergency fund, positive cash flow, core insurance in place
- Stage 2: Eliminate high-cost debt
- Stage 3: Capture free money — employer match, HSA, tax-advantaged accounts
- Stage 4: Invest intentionally toward each specific goal
- Stage 5: Optimize — tax efficiency, asset allocation, estate planning
- Stage 6: Build and protect long-term wealth

---

## Section 5 - Estate Planning Checklist
- List what this client currently has vs. what they need for their life stage
- For anything missing, state in one line why it matters
- Minimum required at their age: will, POA (financial + medical), 
  beneficiary designations confirmed on all accounts
- If they have dependents or significant assets, address trust considerations

## Section 6 - Advanced Concepts Worth Understanding
3 topics relevant to where this client is headed — even if not 
immediately actionable. Calibrated for a financially sophisticated reader.
Only include concepts plausibly relevant to this client's trajectory.
Examples: Roth conversion ladders, backdoor Roth, asset location strategy,
tax-loss harvesting, sequence-of-returns risk, mega backdoor Roth, 
IRMAA thresholds, I-bonds, QCD strategies.

---

# Section 7 - Goals, Implementation & Monitoring

## Concrete Goals
For each of the client's 1-3 primary goals state:
- Target amount in dollars
- Realistic timeline
- Monthly savings required to hit it
- Recommended account or vehicle
- Investment risk appropriate to the time horizon

## Priority Action Plan

*Next 30 days:*
- 3 specific, immediately actionable tasks

*Next 3-12 months:*
- 3 medium-term milestones

*12-24 months:*
- 3 longer-term targets to reach or review

## Monitoring Checkpoints
- Annual review: 4 specific things to check each year 
  (savings rate, debt progress, insurance adequacy, net worth, tax accuracy)
- Trigger events: 3 life events that should prompt an immediate plan review
- Monthly benchmark: one specific number the client tracks each month 
  to confirm they are on track

---

# Summary Table

| Focus Area | Current Situation | Recommended Action | Estimated Impact |
|---|---|---|---|
| Emergency Fund | | | |
| Debt | | | |
| Retirement Savings | | | |
| Tax Strategy | | | |
| Insurance | | | |
| Estate Planning | | | |
| Goals | | | |

Every cell must be specific to this client — no placeholder text.

---

IMPORTANT RULES:
- Every dollar amount must be grounded in the user's actual income figure.
- Every % must be translated into a dollar amount at their income level.
- Tone: warm, direct, professional. Like a trusted advisor in plain English — not a textbook.
- Do NOT say "consult a financial advisor" as a cop-out. You ARE the advisor.
  You may say "a CFP can help implement this" once at the very end if appropriate.
- Use web search for current IRS limits, rates, and program rules. Always cite with (Source: [Title](URL)).
- If data is missing, state your assumption clearly, then give the best guidance you can.
\end{verbatim}
\end{tcolorbox}

\subsection{Trend and Risk Evaluation Prompt}\label{trend-risk-prompt}
\noindent
\textbf{Functionality Overview:}

\begin{enumerate}[label=,leftmargin=2.23em,labelsep=0.2mm]

        \item[--] {One prompt interprets financial ratios and volatility metrics to assess an asset’s overall trend direction (e.g., increasing or stable) and risk level (e.g., moderate or high).}
        \begin{itemize}
            \item[*] {The \textbf{Trend \& Risk LLM} uses GPT-4 to evaluate parameters such as beta, P/E ratio, debt-to-equity, and recent price movement to assign both trend and risk categories.}
        \end{itemize}
\end{enumerate}

\noindent
\textbf{Input Variables:}
\begin{enumerate}[label=,leftmargin=2.23em,labelsep=0.2mm]

        \item[--] data (str): A text representation of financial metadata extracted from the asset’s yFinance profile, including volatility, beta, P/E, and debt ratios.
\end{enumerate}

\noindent
\textbf{Output Variables:}
\begin{enumerate}[label=,leftmargin=2.23em,labelsep=0.2mm]

        \item[--] trend\_risk (str): A concise paragraph summarizing both the price trend and risk classification.
\end{enumerate}

\noindent
\textbf{LLM Settings---Trend \& Risk Evaluation:}
\begin{enumerate}[label=,leftmargin=2.23em,labelsep=0.2mm]

        \item[--] \texttt{gpt-4-turbo}
        \item[--] Prompt Strategy: few-shot learning    
        \item[--] temperature = 0.2
        \item[--] top\_p = 0.5
\end{enumerate}

\noindent
\textbf{Prompt---Trend \& Risk Evaluation:}
\footnotesize
\begin{tcolorbox}[
    breakable,
    width=1.1\linewidth,
    left=4mm, right=4mm  % adjust so text doesn't touch edges
]
\begin{verbatim}
You are an expert in assessing both the overall trend and risk of a 
stock based on financial metrics and ratios. Based on the information 
provided, determine if:
- The overall trend is "increasing," "decreasing," or "stable."
- The overall risk level is "Low," "Moderate," or "High."
- Provide

Here are some examples:

Example 1:
User provides the following data:
- Volatility: 1.5
- Debt-to-Equity Ratio: 0.8
- Price-to-Earnings Ratio (P/E): 20
- Beta: 1.2
- Recent Trend: increasing
AI: The stock is showing an increasing trend, with its price consistently 
rising over the past year. This suggests growing investor confidence and 
potential for continued growth. The risk level is moderate due to the 
higher beta of 1.2, indicating the stock is slightly more sensitive to 
market fluctuations than average. However, the moderate P/E ratio of 20 
reflects a valuation that is reasonable for its earnings, and the low 
debt-to-equity ratio of 0.8 indicates strong financial health, which helps 
offset some of the market-related risks. Overall, this stock may appeal to 
investors seeking balanced growth opportunities with manageable risk.

Example 2:
User provides the following data:
- Volatility: 2.0
- Debt-to-Equity Ratio: 1.5
- Price-to-Earnings Ratio (P/E): 40
- Beta: 1.6
- Recent Trend: decreasing
AI:  The stock is currently on a decreasing trend, with its price declining 
over the past year, which could indicate waning investor confidence or 
broader market pressures. The risk is high, driven by multiple factors: 
the elevated beta of 1.6 reflects significant sensitivity to market 
movements, while the high debt-to-equity ratio of 1.5 suggests the company 
is heavily leveraged, potentially increasing financial vulnerability. 
Additionally, the high P/E ratio of 40 points to possible overvaluation, 
meaning the stock's price may not be well-supported by its earnings. This 
combination of factors makes the stock more suitable for investors with a 
high risk tolerance who are prepared for potential volatility and downturns.

Example 3:
User provides the following data:
- Volatility: 0.7
- Debt-to-Equity Ratio: 0.4
- Price-to-Earnings Ratio: 15
- Beta: 0.9
- Recent Trend: stable
AI: The stock is exhibiting a stable trend, with its price remaining 
relatively consistent over the past year. This indicates a steady 
performance and low market-driven volatility. The risk level is low, 
supported by the stock's beta of 0.9, which reflects below-average 
sensitivity to market changes. Additionally, the low debt-to-equity 
ratio of 0.4 highlights the company’s strong financial position, while 
the modest P/E ratio of 15 suggests the stock is reasonably valued 
relative to its earnings. This stability and financial prudence make 
the stock a suitable option for conservative investors seeking reliable 
returns with minimal risk exposure.

Evaluate the data and return the trend and risk as a concise output.
\end{verbatim}
\end{tcolorbox}

\subsection{Article-Based Stock Analysis Prompt}\label{article-synthesis-prompt}
\noindent
\textbf{Functionality Overview:}

\begin{enumerate}[label=,leftmargin=2.23em,labelsep=0.2mm]

        \item[--] {One prompt synthesizes recent financial news about a stock, mutual fund, or ETF to assess sentiment and extract relevant investment signals. It uses scraped articles to provide a market overview and highlight performance drivers.}
        \begin{itemize}
            \item[*] {The \textbf{Article Synthesis LLM} combines multiple sources into a coherent response, identifying both positive and negative trends such as earnings growth, volatility, competition, and regulatory risk.}
        \end{itemize}
\end{enumerate}

\noindent
\textbf{Input Variables:}
\begin{enumerate}[label=,leftmargin=2.23em,labelsep=0.2mm]

        \item[--] articles (list): A list of dictionaries, each containing a scraped news article’s title, URL, and paragraph content.
        \item[--] question (str): A user-submitted question related to a specific asset, or a default system prompt if none is provided.
\end{enumerate}

\noindent
\textbf{Output Variables:}
\begin{enumerate}[label=,leftmargin=2.23em,labelsep=0.2mm]

        \item[--] stock\_analysis (str): A structured response summarizing financial sentiment, performance drivers, and risk factors.
\end{enumerate}

\noindent
\textbf{LLM Settings---Article Synthesis:}
\begin{enumerate}[label=,leftmargin=2.23em,labelsep=0.2mm]

        \item[--] \texttt{gpt-4-turbo}
        \item[--] Prompt Strategy: few-shot learning    
        \item[--] temperature = 0.5
        \item[--] top\_p = 1
\end{enumerate}

\noindent
\textbf{Prompt---Article Synthesis:}
\footnotesize
\begin{tcolorbox}[
    breakable,
    width=1.1\linewidth,
    left=4mm, right=4mm  % adjust so text doesn't touch edges
]
\begin{verbatim}
You are an expert financial analyst. Based on the articles and data 
provided, synthesize an objective overview of the specific stock, mutual 
fund, or ETF in question. Highlight both positive and negative indicators, 
including factors like price history, sales and profit trends, and sector 
performance. Provide reasons supporting each point, and summarize key 
information from the articles. Include citations when referring to 
specific claims.

Here are some examples:

User: Is Tesla (TSLA) a good investment?
AI:
Positive Indicators for TSLA:
Strong Sales Growth: Tesla reported a 50% growth in sales year-over-year, 
indicating increasing demand for its electric vehicles. (Source: Tesla's 
Q2 Earnings Report)
Expansion of Production Capacity: New factory openings in Berlin and 
Texas are expected to boost production and reduce delivery times.
Market Leadership: Tesla continues to lead the EV market, maintaining a 
significant market share despite emerging competitors.
Technological Advancements: Ongoing investments in battery technology and 
autonomous driving capabilities.
Negative Indicators for TSLA:
Rising Competition: Increased competition from traditional automakers 
entering the EV market may affect Tesla's market share.
Stock Price Volatility: Recent fluctuations in stock price could indicate 
market uncertainty.
Regulatory Risks: Potential regulatory challenges in key markets like 
China and Europe.
Supply Chain Constraints: Global semiconductor shortages may impact 
production timelines.

User: Should I invest in Vanguard 500 Index Fund (VFIAX)?
AI:
Positive Aspects of VFIAX:
Consistent Performance: VFIAX has historically tracked the S&P 500 closely, 
offering steady returns over the past decade.
Low Expense Ratio: The fund offers a low-cost way to invest in a broad 
market index.
Investor Confidence: Vanguard funds have seen record inflows this year, 
indicating strong investor trust. (Source: Vanguard Annual Report 2023)
Diversification: Provides exposure to 500 large-cap U.S. companies, 
reducing company-specific risk.
Negative Aspects of VFIAX:
Market Risk Exposure: Being tied to the S&P 500, the fund is susceptible 
to overall market downturns.
Limited International Exposure: Concentration in U.S. markets may miss out 
on global growth opportunities.
Economic Uncertainty: Factors like inflation and interest rate changes 
could impact future performance.
Dividend Yield: Lower dividend yields compared to some other investment 
options.
\end{verbatim}
\end{tcolorbox}

\section[\appendixname~\thesection]{Backend Implementation }\label{backend-appendix}
The backend is implemented using FastAPI and exposes two primary POST endpoints: \texttt{/query} and \texttt{/questionnaire}, each responsible for handling different branches of the user workflow.

The \texttt{/query} endpoint is used to generate both general financial insight and single-stock analysis, and always invokes the \texttt{run\_program()} function. Within \texttt{run\_program()}, conditional logic determines which analyses are executed. If the \texttt{symbol} field (denoted as \texttt{ticker\_sym} in the code) is not equal to \texttt{'none'}, the system retrieves quantitative stock metrics and recent news using \texttt{yfinance} and invokes GPT-4 to generate qualitative stock summaries, trend assessments, and risk interpretations. If the \texttt{questionGeneral} field (denoted as \texttt{user\_question\_general}) is not equal to \texttt{'none'}, the system calls \texttt{general\_insights()}, which formats the user’s questionnaire responses into a structured prompt and queries the Perplexity API (\texttt{sonar-pro} model) for structured general financial insight. Both branches may be executed within the same request, enabling combined general and stock-level analysis.

The \texttt{/questionnaire} endpoint supports questionnaire-driven workflows, including the Stock Insights Questionnaire. When the user provides both a risk tolerance and one or more selected sectors, the endpoint invokes the \texttt{classify\_risk\_sector()} function. This function first retrieves the top 50 companies for each selected sector using the \texttt{yfinance} sector API and downloads six months of historical daily price data. For each ticker, the system computes historical volatility, recent price change, dividend yield, return on capital, and earnings yield, and derives a combined Magic Formula rank based on return-on-capital and earnings-yield rankings.

Each stock is assigned to one of five volatility-based risk categories—Very Conservative, Conservative, Moderate, Aggressive, or Very Aggressive—according to fixed volatility thresholds. The results are returned as a structured dictionary mapping risk categories to lists of qualifying stocks along with their associated metrics. If the user submits the Stock Insights Questionnaire without specifying both a risk tolerance and at least one sector, the back end returns the form payload without generating sector-level stock tables; in practice, the Stock Insights workflow assumes that users select at least one sector in order to receive categorized stock results.

\subsection{General Insight Generation via Perplexity}

When users select the \texttt{General Insight} workflow or submit a query under \texttt{mode=both}, the back end routes the request through the \texttt{/query} endpoint. If the submitted payload contains a non-empty \texttt{questionGeneral} field, the back end invokes the \texttt{general\_insights()} function. This function synthesizes a structured prompt by concatenating the user’s responses to the financial questionnaire—age group, employment status, debt, savings, investment goals, risk tolerance, and planning challenges—using labeled text formatting. If the user has not entered a specific financial query, a default fallback question (“Please give me general insight.”) is inserted automatically. The fully composed prompt is passed to the Perplexity API, using the Sonar-Pro model. 

The system prompt instructs Perplexity to return structured strategic financial insights with the following format: Key Findings, Action Options, Cost/Benefit Analysis, Strategic Considerations, and a list of Reference URLs. The model response is parsed as plain text and split into these constituent blocks using hardcoded section headers. The parsed insight is returned to the front end in JSON format under the \texttt{insightText} and \texttt{insightUrls} fields. 

When running in \texttt{mode=both}, the system combines the Perplexity-generated general insight with the stock evaluations and then shows them together on the summary page. %\textcolor{blue}{FIXED} \textcolor{red}{John Notes: This sentence sounds way to complex, please simplify the language and make it more clear.} 
This pathway tries to ensure that user context is faithfully encoded into the LLM prompt, and that the output remains structured and interpretable by both the front end and end user.

\subsection{Sector-Level Investment Classification}

When the user selects the \texttt{Stock Insight} workflow, the front end sends their selected sectors and risk tolerance to the \texttt{/questionnaire} endpoint. The back end calls the \texttt{classify\_risk\_sector()} function, which matches the user’s risk preference to specific volatility categories. Choosing \texttt{safeAssets} means the user is limited to conservative options (very conservative, conservative and moderate). Selecting \texttt{riskyAssets} gives the user only higher-volatility choices (moderate, aggressive, and very aggresive). If the user selects \texttt{both}, the system considers the full range of categories. 
% \textcolor{blue}{FIXED} \textcolor{red}{John Notes: The language here can be simplified, I think we would need to explain what it means to be safeAssets, RiskyAssets or both, safeAssets means that the user queues for only very concervative, convervative and moderate, RiskyAssets will provide the user with only moderate, aggressive and very aggressive investment vehicles that have high stock volatility}. 

For each selected sector, the function pulls a list of companies using the predefined \texttt{sector\_to\_companies} dictionary. 

Then, for each ticker, it retrieves 6-month historical price data via the \texttt{yfinance} API and computes three metrics: historical volatility (standard deviation of percentage price changes), overall price change, and dividend yield.

Volatility values are then used to assign each company into one of five discrete risk categories: Very Conservative, Conservative, Moderate, Aggressive, or Very Aggressive, based on hardcoded threshold values. These buckets allow the system to match individual stocks to the user’s stated risk preference. The output for each sector is a dictionary mapping risk categories to lists of qualifying stock tickers. These results are combined across all sectors and returned to the front end as a JSON object. The front end renders this as a spreadsheet-style interactive table, with each ticker linked to the single-stock analysis view. This pathway forms the core of Algorithm \ref{sector-based-alg} and enables targeted investment exploration based on both user intent and historical price dynamics.

\subsection{Single Stock Evaluation via GPT}

When a user inputs a specific stock ticker or clicks a company name from the sector spreadsheet interface, the front end triggers a POST request to the \texttt{/query} endpoint with the \texttt{symbol} field populated. The back end then executes the \texttt{run\_program()} function, which handles stock-specific data aggregation and GPT-powered synthesis. First, it uses \texttt{yfinance} to fetch the historical price data for the past six months. It then runs \texttt{get\_price\_change\_volatility\_dividend()}, which calculates the asset’s price change percentage, historical volatility (standard deviation of returns), and average dividend yield. These metrics are formatted as a string and later analyzed by GPT.

Next, \texttt{fetch\_yfinance\_urls()} extracts recent news article URLs from the ticker’s news metadata. These links are parsed by \texttt{scrape\_urls\_content()}, which uses \texttt{BeautifulSoup} to extract and truncate article content to manageable input size. The resulting corpus is passed to \texttt{article\_output\_synthesis()}, which sends a prompt to OpenAI’s \texttt{gpt-4-turbo} model to summarize news themes and respond to the user’s original stock-related question. Separately, the numerical stock metrics string is fed to \texttt{determine\_trend\_and\_risk()}, which prompts GPT-4 to assess recent performance and implied volatility in qualitative terms.

The outputs of both GPT prompts are concatenated to form a comprehensive stock analysis, which is returned to the front end along with a dictionary of referenced article titles and URLs. The final result is provided in the summary page. This pathway constitutes Algorithm 3 and enables qualitative interpretation of financial signals and news narratives for any stock ticker.

\subsection{PDF Report Generation}

\ourapproach provides built-in PDF export functionality that allows users to download summaries of their financial insights and stock classifications. This feature is implemented entirely on the front end using the \texttt{jsPDF} and \texttt{autotable} JavaScript libraries. In \texttt{summary\_page.html}, a button labeled “Download PDF” triggers a function that initializes a new \texttt{jsPDF} document, injects section headers such as \texttt{General Insight}, \texttt{Trend and Risk}, and \texttt{News Summary}, and appends the corresponding textual content retrieved from \texttt{sessionStorage} (keys like \texttt{insightText} and \texttt{insightUrls}). The PDF is then automatically downloaded with a timestamped filename.

In \texttt{spreadSheet.html}, a similar export workflow is implemented for sector classification results. When the user clicks the “Download PDF” button, the front end parses the hierarchical risk-bucketed dictionary of classified stocks from \texttt{sessionStorage}, formats the content into a tabular structure grouped by sector and risk level, and uses \texttt{autotable} to render the content in the PDF. Each ticker symbol, sector name, and classification label is included in the final document. The PDF files are generated entirely client-side, without additional back end processing, ensuring fast and private data handling. This feature supports both single-mode (sector or insight) and combined workflows, enabling users to export all relevant findings in a consolidated offline format.

\section[\appendixname~\thesection]{Quantitative Risk Evaluation} \label{accuracy-analysis}

The data-driven evaluation measures how well  the volatility-based classifications of \ourapproach in one six-month period correlate with the volatility of the same stocks in the following six-month period. Intuitively, low risk companies should continue to be low risk (and often low reward) and conversely for high risk companies. That is in fact what we find.

%\subsection{Evaluation Design and Rationale} To create a benchmark for evaluating the consistency of \ourapproach's risk classification framework, we created a dataset using historical volatility data from the technology sector across ten defined six month intervals. 

\subsubsection{Data Setup}

For each six-month categorization period $P$, \ourapproach assigns stocks to one of five risk categories based on their observed historical volatility, computed as the standard deviation of daily closing prices over the preceding six months. Here, $\sigma$ denotes the historical return volatility, which is computed as the standard deviation of daily percentage return for each stock over the preceding six-month window. 
The volatility thresholds are fixed and consistent across all periods: Very Conservative ($0 \leq \sigma \leq 5$), Conservative ($5 < \sigma \leq 10$), Moderate ($10 < \sigma \leq 15$), Aggressive ($15 < \sigma \leq 20$), and Very Aggressive ($\sigma > 20$), as implemented in the \ourapproach classification engine. 

Figure~\ref{tab:volatility-backtest} visualizes this categorization by plotting, for each risk group, the average volatility of all stocks assigned to that group during period $P$. Each bar therefore represents an aggregate volatility level produced by the classification rules, rather than individual stock outcomes. The corresponding values for the subsequent six-month window ($P+1$) are shown to evaluate whether stocks assigned to lower- or higher-risk buckets in period $P$ continue to exhibit similar relative volatility behavior in the following period.

This benchmark was structured to allow a direct comparison between the average volatility of stocks within each risk category during period $P$ and the observed average volatility during period $P+1$. This design enables an assessment of whether \ourapproach’s volatility-based classification remains informative across adjacent time horizons.

The benchmark dataset includes:
\begin{itemize}
    \item Ten 6-month time periods between July 2020 and June 2025.
    \item Five volatility-based risk groups per period: Very Conservative, Conservative, Moderate, Aggressive, and Very Aggressive.
    \item Matched average volatility values for each group in both period $P$ and period $P+1$.
\end{itemize}
% This approach allows us to assess \ourapproach's performance using a quantitative, correlation-based framework.

\subsubsection{Consistency Measurement Method}
To evaluate the consistency between the classification periods and the future average volatilities, we computed the linear correlation coefficient (r) between:
\begin{itemize}
    \item Volatility during Period $P$ of the five groups
    \item Volatility of the same stock groups during period $P+1$.
\end{itemize}

For example, for the first half of 2021 ($P=\,$2021-01-01 to 2021-06-30), the five category averages are
\[
[\,3.3,\; 7.1,\; 13.1,\; 17.1,\; 30.7\,],
\]
and for the second half of 2021 ($P{+}1=\,$2021-07-01 to 2021-12-31) they are
\[
[\,3.2,\; 7.2,\; 12.1,\; 17.0,\; 33.2\,].
\]

The correlation coefficient (r) measures how strongly average volatility in a time period $P$ predicts the volatility in the next six month period $P+1$ for the same stocks. A value closer to 1 indicates a strong positive consistency, which means that \ourapproach's classification in period $P$ is  predictive of volatility for those same stocks in period $P+1$.

Using a non-parametric statistical test \cite{shasha2010statistics} to avoid the need for statistical distribution assumptions, we compute the correlation between the above five-dimensional vectors to be
$r_{\text{category}}=0.997$  having a 90\% confidence interval of $[0.973,\; 0.9997]$.
At the stock level for the same pair (across the common tickers in both periods P and P+1), the correlation based on the data of  Table~\ref{tab:volatility-backtest} is
$
r_{\text{stock}}=0.773$ with a  p-value of   $p<0.0001 \; (B=10{,}000).
$
For the next pair (second half of 2021 vs.\ first half of 2022), the category averages
\[
[\,3.2,\; 7.2,\; 12.1,\; 17.0,\; 33.2\,] \;\text{and}\; [\,3.8,\; 7.4,\; 12.0,\; 17.3,\; 38.1\,]
\]
yield a correlation
$
r_{\text{category}}=0.997$ with a 90\% confidence interval of $[0.969,\; 0.9997]$,
while the stock-level correlation in Table~\ref{tab:volatility-backtest} is
$
r_{\text{stock}}=0.915$ with p-value $p<0.0001$.

 To further assess the reliability of each of the correlation scores, we computed a 90\%\ confidence intervals using bootstrapped error estimates.

The linear correlation coefficients for each of the ten six-month windows ranged from 0.9924 to 0.9991, with every interval showing positive correlation between past and future average volatility. These findings show that \ourapproach's risk classification framework is predictive of future volatility behavior. Admittedly, this is not very surprising, but the level of correlation surprised us. Stock picking is not entirely random: a volatile (respectively, stable) stock now will likely remain volatile (respectively, stable) six months from now.   

Table \ref{tab:volatility-backtest}
 shows a bar chart visualizing the correlation values for all tested time periods. The complete underlying data—including the volatility averages for each risk category in $P$ and $P+1$ is shown in Table \ref{tab:volatility-backtest}.

\begin{figure}[H] \includegraphics[width=1\textwidth]{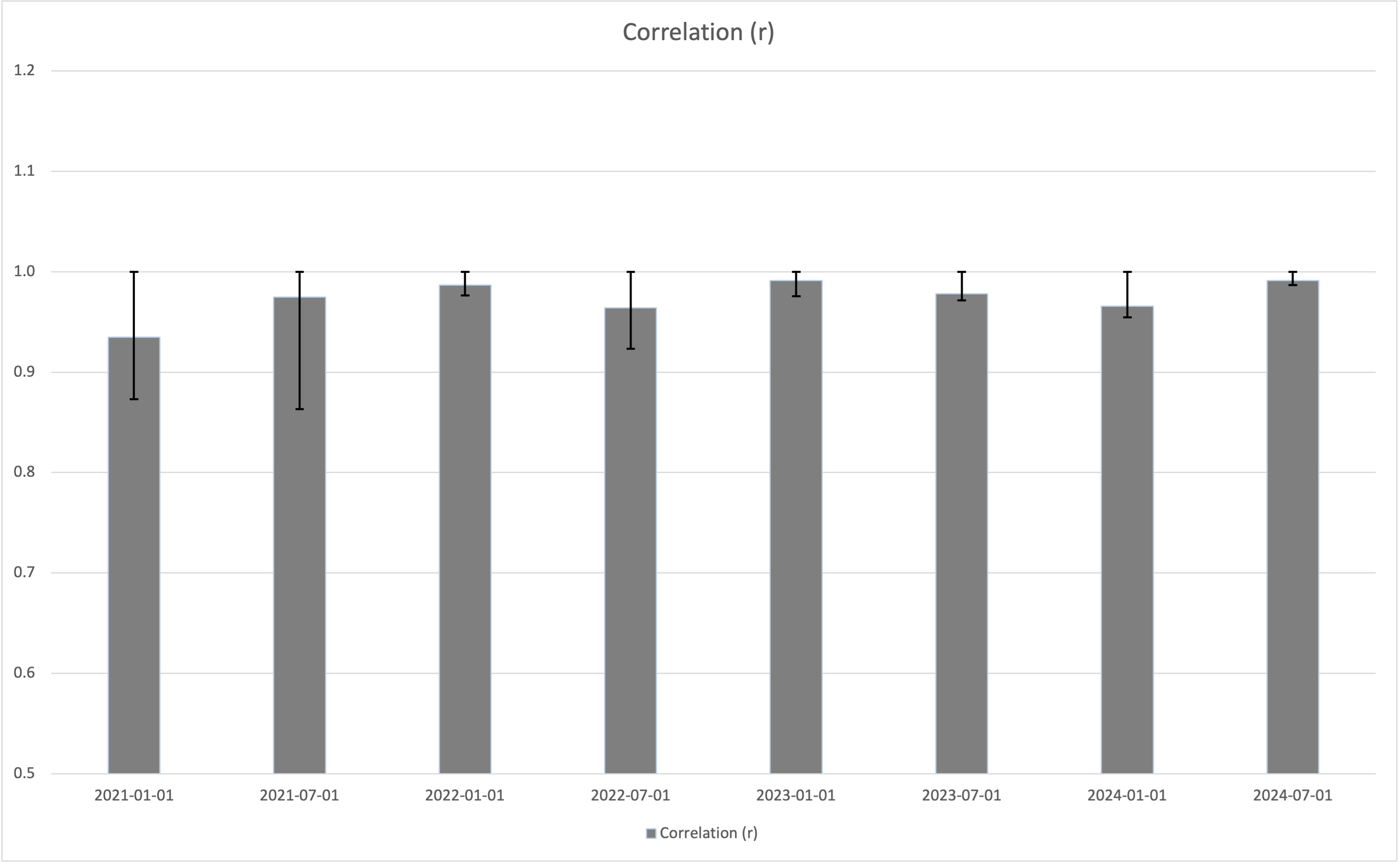}
    \caption{Correlation  between average volatility in period $P$ and $P+1$ across ten consecutive six-month intervals from July 2020 to June 2025. Each point represents the linear correlation of the five average volatilities of each risk category at period $P$ (e.g., first half of 2021)  with the volatilities of the same sets of stocks in period $P+1$ (e.g., second half of 2021). The error bars represent bootstrapped 90\%\ confidence intervals. \cite{shasha2010statistics}}
\label{fig:CorrelationInvestorNerd}
\end{figure}

\begin{table}[H]
    \centering
    \begin{adjustbox}{width=\textwidth}
    \begin{tabular}{llllll}
    \toprule
    \textbf{P} & \textbf{P+1} & \textbf{Average vals P} & \textbf{Average vals P+1} & \textbf{Correlation r} & \textbf{P-value} \\
    \midrule
   % -- 2020-01-01 & 2020-07-01 & 2.9 7.3 11.5 17.3 28.3 & 2.7 7.6 11.7 17.7 28.8 & 0.822315 & $<0.0001$ \\
    2020-07-01 & 2021-01-01 & 2.7 7.6 11.7 17.7 28.8 & 3.3 7.1 13.1 17.1 30.7 & 0.846618 & $<0.0001$ \\
    2021-01-01 & 2021-07-01 & 3.3 7.1 13.1 17.1 30.7 & 3.2 7.2 12.1 17.0 33.2 & 0.773347 & $<0.0001$ \\
    2021-07-01 & 2022-01-01 & 3.2 7.2 12.1 17.0 33.2 & 3.8 7.4 12.0 17.3 38.1 & 0.915305 & $<0.0001$ \\
    2022-01-01 & 2022-07-01 & 3.8 7.4 12.0 17.3 38.1 & 3.3 7.3 12.4 17.1 36.6 & 0.753808 & $<0.0001$ \\
    2022-07-01 & 2023-01-01 & 3.3 7.3 12.4 17.1 36.6 & 3.3 7.2 12.0 17.1 37.0 & 0.834358 & $<0.0001$ \\
    2023-01-01 & 2023-07-01 & 3.3 7.2 12.0 17.1 37.0 & 3.1 7.4 11.8 17.1 36.1 & 0.902431 & $<0.0001$ \\
    2023-07-01 & 2024-01-01 & 3.1 7.4 11.8 17.1 36.1 & 3.0 6.6 12.4 16.8 32.6 & 0.755046 & $<0.0001$ \\
    2024-01-01 & 2024-07-01 & 3.0 6.6 12.4 16.8 32.6 & 4.1 8.2 12.8 17.4 44.9 & 0.743741 & $<0.0001$ \\
    2024-07-01 & 2025-01-01 & 4.1 8.2 12.8 17.4 44.9 & 2.8 6.8 12.5 17.1 38.3 & 0.814570 & $<0.0001$ \\
    \bottomrule
    \end{tabular}
    \end{adjustbox}
    \caption{Each row compares two adjacent periods P and P+1,  reporting the five risk-category averages in P and P+1 for ten consecutive half-years (July 2020–Jun 2025), the correlation of stock-level volatilities between the two periods, and the permutation-test p-value for significance.}
    \label{tab:volatility-backtest}
\end{table}

\section[\appendixname~\thesection]{InvestorNerd Questionnaire Inputs}
\label{app:questionnaire}
The following pages reproduce the  InvestorNerd questionnaire as completed 
for this experiment. The selected responses, highlighted in green, represent 
the household profile used to generate the 
report  in Section~\ref{sec:comparison}.

\includepdf[pages=-]{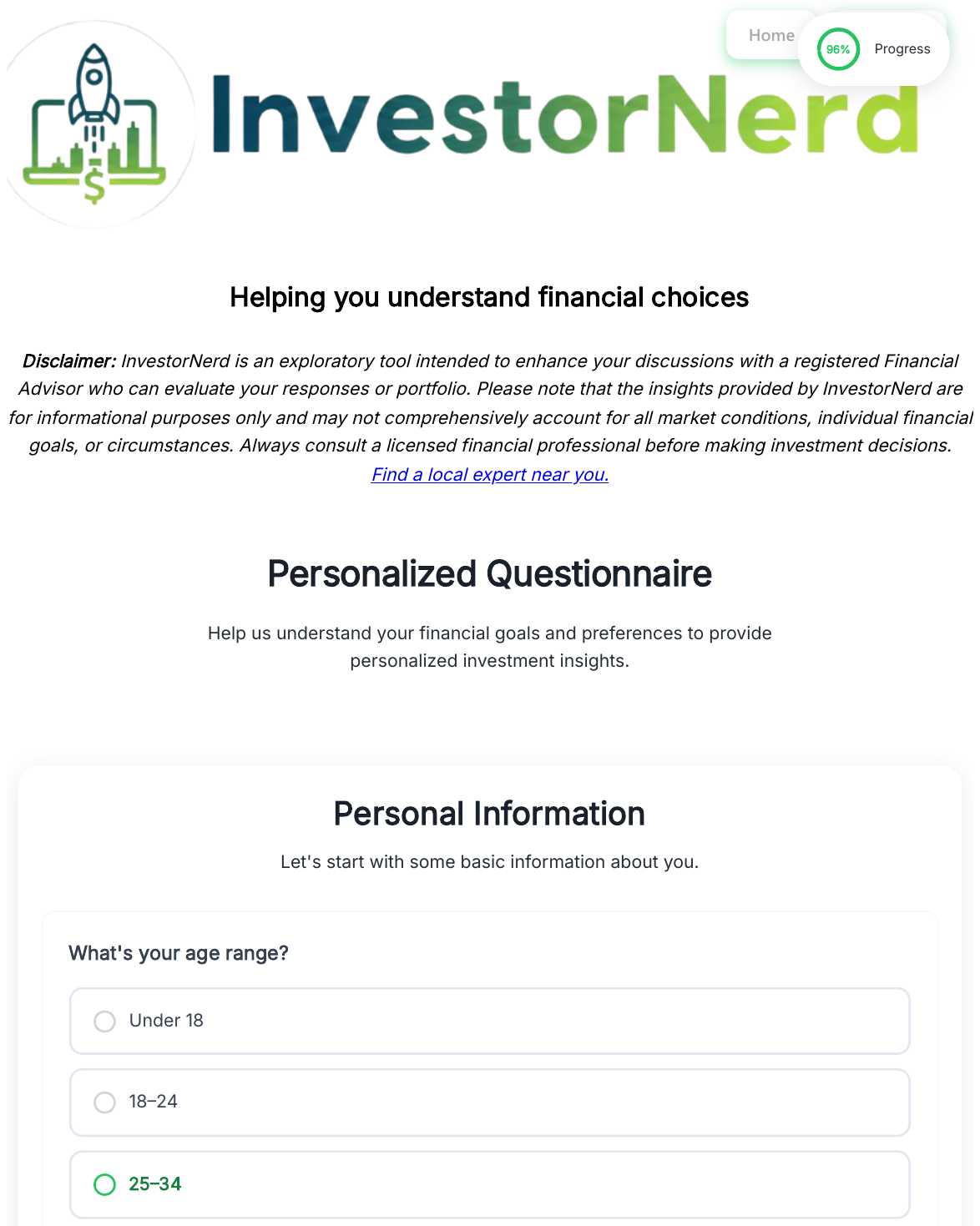}

\begin{comment}
\section[\appendixname~\thesection]{InvestorNerd Report - Version 1}
\label{app:report_v1}
The following pages reproduce the InvestorNerd report generated by the 
first version of the tool, using the questionnaire inputs described in 
Appendix~\ref{app:questionnaire}. This report formed the basis of the 
initial evaluation referenced in Section~\ref{sec:comparison}.

\includepdf[pages=-]{InvestorNerd_Full_Report_29_1.pdf}
\end{comment}
\section[\appendixname~\thesection]{InvestorNerd Report }
\label{app:report_v2}
The following pages show the InvestorNerd report  using the  questionnaire inputs 
described in Appendix~\ref{app:questionnaire}. This report forms the 
primary basis of the comparative analysis conducted in 
Section~\ref{sec:comparison}.

\includepdf[pages=-]{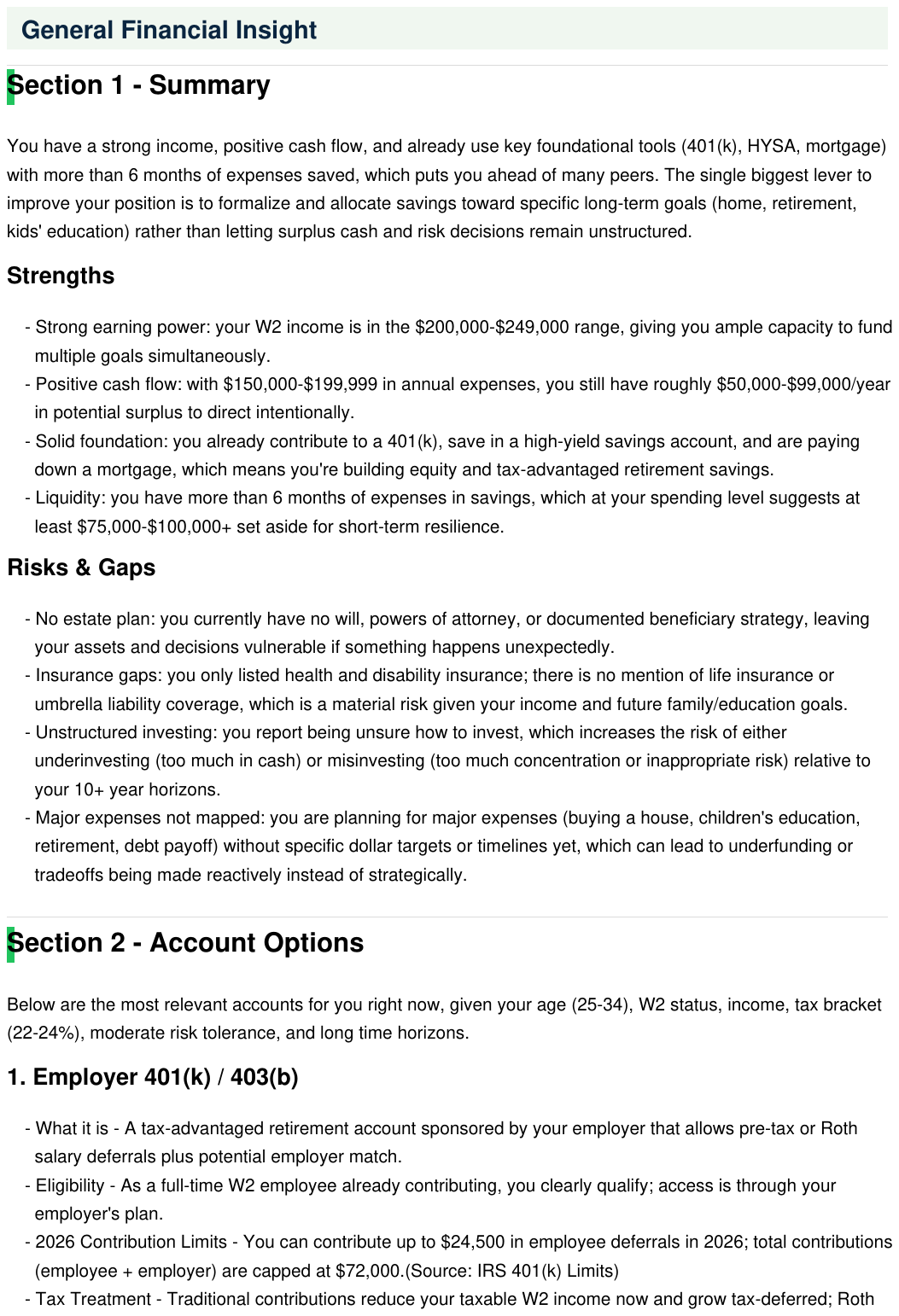}

\end{document}